 \documentclass[final,3p,times]{elsarticle}
\usepackage{amssymb}
\usepackage{amsmath}
\usepackage{tensor}
\usepackage{color}
\usepackage[colorlinks=true,linkcolor=blue,citecolor=blue,urlcolor=blue]{hyperref}
\newcommand{\bs}{\begin{split}}
\newcommand{\es}{\end{split}}

\begin{document}

\begin{frontmatter}

%% Title, authors and addresses

%% use the tnoteref command within \title for footnotes;
%% use the tnotetext command for theassociated footnote;
%% use the fnref command within \author or \affiliation for footnotes;
%% use the fntext command for theassociated footnote;
%% use the corref command within \author for corresponding author footnotes;
%% use the cortext command for theassociated footnote;
%% use the ead command for the email address,
%% and the form \ead[url] for the home page:
%% \title{Title\tnoteref{label1}}
%% \tnotetext[label1]{}
%% \author{Name\corref{cor1}\fnref{label2}}
%% \ead{email address}
%% \ead[url]{home page}
%% \fntext[label2]{}
%% \cortext[cor1]{}
%% \affiliation{organization={},
%%             addressline={},
%%             city={},
%%             postcode={},
%%             state={},
%%             country={}}
%% \fntext[label3]{}

\title{Energy extraction through magnetic reconnection from a Kerr-Newman black hole in perfect fluid dark matter}

%% use optional labels to link authors explicitly to addresses:
 \author[GC,WPI,CfA]{Shanshan Rodriguez\corref{cor1}}
 \ead{rodriguezs@grinnell.edu}
 \cortext[cor1]{Corresponding Author}
  %--------------------------------------
  \author[GC]{Alex Sidler}
 \ead{sidleral@grinnell.edu}
 %--------------------------------------
   \author[GC,WPI]{Leo Rodriguez}
 \ead{rodriguezl@grinnell.edu}
 %--------------------------------------
    \author[WPI]{L. R. Ram-Mohan}
 \ead{lrram@wpi.edu}
 %--------------------------------------
\affiliation[GC]{organization={Department of Physics, Grinnell College},
%             addressline={},
             city={Grinnell IA},
             postcode={50112},
%             state={IA},
             country={USA}}
%--------------------------------------
\affiliation[WPI]{organization={Department of Physics, Worcester Polytechnic Institute},
%             addressline={},
             city={Worcester MA},
             postcode={01609},
%             state={},
             country={USA}}

\affiliation[CfA]{organization={Harvard-Smithsonian Center for Astrophysics},
%             addressline={},
             city={Cambridge MA},
             postcode={02138},
%             state={},
             country={USA}}
%\author{} %% Author name

%% Author affiliation
%\affiliation{organization={},%Department and Organization
%            addressline={}, 
%            city={},
%            postcode={}, 
%            state={},
%            country={}}

%% Abstract
\begin{abstract}
In this work, we provide a thorough analysis of energy extraction via magnetic reconnection, a novel mechanism recently proposed by Comisso and Asenjo, for a Kerr-Newman black hole immersed in a perfect fluid dark matter (PFDM) background. Our studies focus on the impact of black hole spin $a$, electric charge $Q$ and PFDM parameter $\lambda$ on the horizons, ergoregion and circular geodesics at the equatorial plane of this black hole, and how they further influence the reconnection efficiency and energy extraction rate. Our results show that the outer horizon and the size of ergoregion do not vary monotonically with increasing dark matter parameters $\lambda$ until reaching its critical value $\lambda_c$ due to the combined counteracting effect between the black hole's charge and dark matter parameter.  We identify the optimal combinations of $a$, $Q$ and $\lambda$ that allow for efficient energy extraction and high extracted power, even when the black hole is not spinning near its extremal limit. Our results ease the stringent conditions observed in other rotating black holes, where achieving comparable levels of extracted power and reconnection efficiency typically requires a near-extremal spin.
\end{abstract}

%%Graphical abstract
%\begin{graphicalabstract}
%\includegraphics{grabs}
%\end{graphicalabstract}

%%Research highlights
%\begin{highlights}
%\item Research highlight 1
%\item Research highlight 2
%\end{highlights}

%% Keywords
\begin{keyword}
Dark matter \sep Black holes \sep Rotating charged \sep Geodesics \sep Energy extraction \sep Magnetic reconnection 

%% PACS codes here, in the form: \PACS code \sep code

%% MSC codes here, in the form: \MSC code \sep code
%% or \MSC[2008] code \sep code (2000 is the default)

\end{keyword}

\end{frontmatter}

%% Add \usepackage{lineno} before \begin{document} and uncomment 
%% following line to enable line numbers
%% \linenumbers

%% main text
%%

%% Use \section commands to start a section
%---------------------------------------
\section{Introduction}\label{Sec:intro}
%---------------------------------------
%% Labels are used to cross-reference an item using \ref command.
Astrophysical black holes, among the most enigmatic and dense objects predicted by Einstein's general theory of relativity, have captivated scientists for their profound influence on high-energy astrophysical phenomena. In recent years, the development of advanced instruments such as gravitational wave interferometers \cite{LIGOScientific:2016aoc, LIGOScientific:2019fpa} and the Event Horizon Telescope (EHT) \cite{EventHorizonTelescope:2019ths, EventHorizonTelescope:2019dse, EventHorizonTelescope:2022wkp, EventHorizonTelescope:2022wok} have significantly enhanced the study of black hole physics, opening new avenues for phenomenological research. One of the primary focuses in high-energy astrophysics is to identify the sources of relativistic jets and beams such as gamma-ray bursts (GRBs). Black holes, as natural candidates capable of driving energy production in the universe, are suggested to be responsible for producing relativistic jets originating from two main sources: the gravitational potential energy of matter falling toward a black hole during accretion, and the electromagnetic field energy generated in the vicinity of a black hole \cite{Misner:1972kx, Penrose:1969pc, Penrose:1971uk}. This area of research continues to be highly valued and actively pursued in the quest to understand how these energetic astrophysical phenomena are connected to the mechanics of black hole energy extraction.

The Penrose process is one of the first well-known frameworks to address energy extraction from a rotating black hole \cite{Penrose:1969pc}. It involves a particle entering the ergosphere of a rotating black hole, and splitting into two parts. A space-like Killing vector inside the ergosphere allows one part to fall into the black hole with negative energy relative to an observer at infinity, while the other escapes with more energy than the original particle, resulting in a decrease in the black hole's total energy and mass. For a rotating black hole with mass $M$ and dimensionless spin $a$, the maximum amount of energy that can be extracted is its rotational energy $E_{tot}=(M-M_{irr})c^2$, where $M_{irr}=M\sqrt{(1+\sqrt{1-a^2})/2}$ is the irreducible mass \cite{Christodoulou:1970wf} and $c$ is the speed of light. For a maximally rotating black hole ($a=1$), it could potentially lose up to 29\% of its mass-energy equivalence (0.29 $Mc^2$). Although theoretically appealing, the Penrose process is less practical in most astrophysical contexts due to the extreme velocities ($>0.5c$) required for the particles involved and its insufficiency in extracting significant rotational energy \cite{Bardeen:1972fi, Wald:1974kya}. Building on Penrose's foundational work, researchers have proposed several alternative energy extraction mechanisms, including the collisional Penrose process \cite{osti_4222462}, superradiant scattering \cite{Teukolsky:1974yv}, the Blandford-Znajek (BZ) process \cite{Blandford:1977ds}, and the magnetohydrodynamic Penrose process \cite{Takahashi:1990bv}. Among these, the BZ process has gained wide acceptance for explaining the energetic jets observed in active galactic nuclei (AGNs) \cite{McKinney:2004ka, Hawley:2005xs, Komissarov:2007rc, Tchekhovskoy:2011zx} and GRBs \cite{Lee:1999se, Tchekhovskoy:2008gq, Komissarov:2009dn}. According to the BZ process, a rotating, electrically neutral black hole in a surrounding magnetic field generated by the accretion disk, induces an electric field that accelerates charged particles in the surrounding plasma. This interaction generates a current, twisting the magnetic field lines into a helical structure, resulting in rotational energy extraction from the black hole that is channeled into powering relativistic jets.

In addition to large-scale powerful jets, frequent bursts of flares originating near the event horizon of accreting black holes have been observed across multiple wavelengths. These include gamma-ray flares in the GeV range from active galactic nuclei (AGN) \cite{Aharonian:2007ig, HESS:2009cfm, Albert:2007zd, Aleksic:2014xsg}, energetic TeV flares from M87* \cite{VERITAS:2010udc, Aliu:2011xm}, and powerful X-ray and infrared flares from Sgr A* \cite{Baganoff:2001kw, Eckart:2004ka, Neilsen:2014kva, GRAVITY:2021hxs}. However, the exact mechanism responsible for triggering these bright flares in the vicinity of black holes remains debated. Koide and Arai \cite{Koide:2008xr} were the first to explore magnetic reconnection as a possible energy extraction mechanism from a rotating black hole. Magnetic reconnection is commonly observed in solar flares and the earth's magnetosphere, where antiparallel magnetic field lines break and reconnect. This mechanism produces thin current sheets and plasmoids that facilitate the rapid conversion of magnetic field energy into kinetic energy carried by released plasma \cite[e.g.][]{Comisso:2016pyg, Comisso:2017arh, PhysRevLett.103.065004, Uzdensky:2014uda, Bhattacharjee2009}. Similar to the Penrose process, magnetic reconnection generates accelerated plasma outflows with equal velocities in opposite directions; one outflow is driven against the black hole’s spin and the other escapes, enabling the energy extraction from the black hole. Multi-frequency radio telescope observations have verified the presence of strong magnetic fields around supermassive black holes, which can influence the dynamics of accretion \cite{Eatough:2013nva}. The EHT collaboration also observed polarized emission in the magnetosphere of the supermassive black hole in M87* \cite{EventHorizonTelescope:2019dse}, confirming twisting magnetic field lines near the horizon that possibly initiate reconnection. General-relativistic MHD simulations \cite[e.g.][]{Ripperda:2021zpn, Ripperda:2020bpz, Parfrey:2018dnc, Komissarov:2005wj, East:2018ayf} suggest that the reconnection process commonly and repeatedly occurs due to the frame-dragging effect in the magnetosphere of rapidly rotating black holes. 

Recently, Comisso and Asenjo extended this mechanism by requiring fast magnetic reconnection within the ergosphere of a rotating black hole and computed the energy extraction rate and reconnection efficiency \cite{Comisso:2020ykg}. They demonstrated that the fast plasma motion facilitated by the reconnection process can extract energy much more efficiently, especially for black holes with high spins and strong magnetic fields, and can potentially surpass the energy extraction efficiency during the BZ process. To distinguish their approach from the earlier work by Koide and Arai, we refer to this as the Comisso-Asenjo (CA) process. Numerous investigations have been conducted for energy extraction via the CA process in a variety of scenarios, such as for Kerr-de Sitter black holes \cite{Wang:2022qmg}, rotating non-Kerr, rotating hairy or regular black holes \cite{Liu:2022qnr, Li:2023htz, Li:2023nmy}, rotating magnetized black holes \cite{Zhang:2024rvk}, rotating black holes with broken Lorentz symmetry \cite{Khodadi:2022dff}, Kerr black holes in modified gravity \cite{ Khodadi:2023juk}, Konoplya-Rezzolla-Zhidenko parametrized black holes \cite{Zhang:2024ptp}, rotating black holes with a U(1) charge \cite{Shaymatov:2023dtt} or a string charge \cite{Carleo:2022qlv}, spinning wormholes \cite{Ye:2023xyv} and Kerr plunging region \cite{Chen:2024ggq}. These studies have revealed differing impacts on energy extraction and plasma energizationf efficiencies by the relevant parameters in their black hole models. However, another critical parameter that has not yet been thoroughly studied is the role of dark matter in the vicinity of a rotating charged black hole during energy extraction via reconnection, as dark matter is also known to have significant impacts on the black hole's spacetime \cite{Hou:2018avu, Haroon:2018ryd}. Indirect observations and the Standard Model of cosmology both suggest that dark matter is abundant and everywhere in the observable universe, constituting at least approximately 27\% of the total matter \cite{Rubin:1980zd, Persic:1995ru, Bertone:2016nfn}. Dark matter is assumed to form halos around black holes in most theories and contributes to the effective black hole mass \cite{Jusufi:2019nrn, Kiselev:2004vy, Navarro:1995iw, Kiselev:2002dx}. Studying black hole models in such realistic astrophysical backgrounds is crucial as dark matter can affect geodesics, leading to significant consequences for the dynamics and energetics of these systems. A series of models to study the interaction between dark matter and black holes have been proposed. One dark matter model type is based on an astrophysical density profile of the dark matter halo that depends on the density parameter and characteristic radius, as dark matter appears to form a density structure near the black hole but disappears asymptotically. This approach is commonly used in galaxy formation and structure simulations, such as the popular NFW, Burkert and Einasto profiles for studying the dark matter halo structure in the cold dark matter (CDM) model \cite{Navarro:1995iw, Navarro:1996gj, Burkert:1995yz, Wang:2019ftp, Xu:2018wow, Xu:2020jpv}. Another alternative dark matter model is to describe dark matter as a perfect fluid \cite{Kiselev:2003ah, Rahaman:2010xs, Li:2012zx}, which has the advantage of providing a well-defined stress-energy tensor leading to an analytical solution, making it easier to study how dark matter interacts gravitationally with black holes, including how it accretes onto the black hole or affects spacetime curvature. Many black hole phenomenological studies have been conducted using the perfect fluid dark matter (PFDM) approach, such as calculating black hole shadow \cite{Hou:2018avu, Haroon:2018ryd}, thin accretion disks \cite{Heydari-Fard:2022xhr}, geodesics \cite{Das:2020yxw, Xu:2017bpz, Rizwan:2018rgs}, thermodynamics \cite{Liang:2023jrj, Rakhimova:2023rie} and superradiance \cite{Liu:2024qso}.

In this paper, we explore energy extraction via magnetic reconnection for a Kerr-Newman black hole in a PFDM background following the framework pioneered by Comisso and Ansenjo. Specifically, we investigate how the presence of dark matter affects the efficiency and dynamics of the energy extraction process. The paper is organized as follows. In section 2, we briefly introduce the Kerr-Newman solution in the presence of PFDM. In section 3, we show the impact of black hole spin $a$, electric charge $Q$, and the PFDM parameter $\lambda$ on the event horizons, ergoregion and circular geodesics of the black hole concerned. In section 4, we introduce and analyze the energy extraction via the CA mechanism in detail by studying the parameter and phase space and computing the extraction rate and reconnection efficiency. We summarize our results in section 5. We have assumed geometrized units $c = G = 1$ throughout this work.

\section{KERR-NEWMAN BLACK HOLE IN PERFECT FLUID DARK MATTER}
Since Kiselev first formulated spherically symmetric black hole metrics in the presence of PFDM \cite{Kiselev:2003ah}, subsequent developments have extended these solutions to include rotation, leading to the Kerr-(A)dS solution \cite{Xu:2017bpz} and the Kerr-Newman solution \cite{Das:2020yxw}. To review the construction of a rotating charged black hole solution in PFDM, we start with the action for Einstein's gravity theory minimally coupled to a U(1) gauge field immersed in perfect fluid dark matter (PFDM), which takes the following form 
\begin{gather}
%\begin{split}
S=\int dx^4 \sqrt{-g}\left\{\frac{R}{16\pi G}-\frac{1}{4}F^{\mu\nu}F_{\mu\nu}+\mathcal{L}_{DM}\right\},\label{eq:action}
%\end{split}.
\end{gather}
where $G$ is the gravitational constant, $g$ is the metric determinant, $R$ is the Ricci scalar, $F_{\mu\nu}=\partial_{\mu}A_{\nu}-\partial_{\nu}A_{\mu}$ is the electromagnetic tensor, and $\mathcal{L}_{DM}$ is the dark matter Lagrangian density. Varying this action with respect to the metric, the Einstein field equations yield
\begin{gather}
%\begin{split}
R_{\mu\nu}-\frac{1}{2}g_{\mu\nu}R=8\pi G\left\{T^M_{\mu\nu}+T^{DM}_{\mu\nu}\right\}=8\pi G T^s_{\mu\nu},\label{eq:EFE}
%\end{split}.
\end{gather}
where $T^s_{\mu\nu}$ represents the total energy-momentum tensor sourcing the black hole in consideration. $T^M_{\mu\nu}$ is the energy-momentum tensor due to the electromagnetic fields generated by the electric charge $Q$
\begin{gather}
T^M_{\mu\nu}=F_{\mu\tau}F\indices{_{\nu}^{\tau}}-\frac{1}{4}g_{\mu\nu}F_{\tau\sigma}F^{\tau\sigma}\\=diag\left\{-\frac{Q^2}{8\pi Gr^4}, -\frac{Q^2}{8\pi Gr^4}, \frac{Q^2}{8\pi Gr^4}, \frac{Q^2}{8\pi Gr^4}\right\}.\label{eq:TM}
\end{gather}
The energy-momentum tensor $T^{DM}_{\mu\nu}$ due to PFDM and its corresponding equation of state are given by
\begin{gather}
T^{DM}_{\mu\nu}=diag\left\{-\rho_m, -\rho_m, P, P\right\},\\\label{eq:TDM}
\frac{P}{\rho_m}=\epsilon-1,
\end{gather}
where $\rho_m$ and $P$ represent the density and pressure of the PFDM, and $\epsilon$ is a constant. A static, spherically symmetric charged black hole solution in PFDM can be obtained 
\begin{gather}
ds^2=-f(r)dt^2+\frac{1}{f(r)}dr^2+r^2d\theta^2+r^2\sin^2\theta d\phi^2,
\end{gather}
where the lapse function $f(r)$ is solved for by requiring $\epsilon=\frac{3}{2}$
\begin{gather}
f(r)=1-\frac{2GM}{r}+\frac{Q^2}{r^2}+\frac{\lambda}{r}\ln\left (\frac{r}{|\lambda |}\right ).
\end{gather}
Here $M$ is the mass of the black hole and we set $M=1$ throughout the paper for simplicity. $\lambda$ is an integration constant of either sign. Due to its connection to the density of the dark matter: $\rho_m=\frac{\lambda}{8\pi G r^3}$, $\lambda$ is also referred to as the PFDM parameter. 

To complete the Kerr-Newman solution in PFDM, the black hole spin $a$ is incorporated using the Newman-Janis algorithm. In Boyer-Lindquist coordinates $(t,r,\theta,\phi)$, where $r$, $\theta$ and $\phi$ represent the radial distance, the polar angle and the azimuthal angle, respectively, the final form of the metric can be described as \cite{Das:2020yxw} 
\begin{gather}
ds^2=g_{tt}dt^2+2g_{t\phi}dtd\phi+g_{rr}dr^2+g_{\theta\theta}d\theta^2+g_{\phi\phi}d\phi^2,
\end{gather}
with
\begin{gather}
g_{tt}=-\frac{1}{\rho^2}\left (\Delta-a^2\sin^2\theta\right ), g_{rr}=\frac{\rho^2}{\Delta}\\
g_{t\phi}=-\frac{a\sin^2\theta}{\rho^2}\left [2Mr-Q^2-\lambda r \ln\left (\frac{r}{|\lambda|}\right )\right], g_{\theta\theta}=\rho^2\\
g_{\phi\phi}=\sin^2\theta \left [r^2+a^2+\frac{a^2\sin^2\theta}{\rho^2}\left (2Mr-Q^2-\lambda r \ln\left (\frac{r}{|\lambda|}\right )\right ) \right ],
\end{gather}
where
\begin{gather}
\Delta=r^2+a^2-2Mr+Q^2+\lambda r\ln \left (\frac{r}{|\lambda|}\right ),\\
\rho^2=r^2+a^2\cos^2\theta.
\end{gather}
This solution reduces to a Kerr-Newman black hole without the presence of PFDM background when the parameter $\lambda=0$.

\section{GEODESICS OF KERR-NEWMAN BLACK HOLE IN PFDM}

Since energy extraction occurs in the ergosphere region of a rotating black hole, it is crucial to first investigate the horizons and stationary limits of a Kerr-Newman black hole in PFDM. Unlike a regular Kerr black hole, where the inner boundary of the ergosphere coincides with the outer event horizon, the structures of horizons and stationary limit surfaces in this context are more complex and interesting. They depend sensitively on the three parameters in the metric: black hole spin $a$, electric charge $Q$, and the PFDM parameter $\lambda$. In the CA mechanism, it is assumed that magnetic reconnection that occurs in the bulk plasma rotating circularly around the black hole is observed at infinity. Thus we solve for the inner ($r_{h-})$ and outer ($r_{h+}$) boundaries of horizons specifically at the equatorial plane $(\theta=\frac{\pi}{2})$ by requiring 
\begin{gather}
\Delta=r^2+a^2-2Mr+Q^2+\lambda r\ln \left (\frac{r}{|\lambda|}\right )=0,\label{eq:Delta}
\end{gather}
and obtain the stationary limit surfaces ($r_{L\pm}$) by setting
\begin{gather}
g_{tt}=-\frac{1}{\rho^2}\left (\Delta-a^2\right )=0.\label{eq:gtt}
\end{gather}
It should be noted here that both positive and negative values of $\lambda$ are permitted and the horizon structure dependence on $\lambda$ with its limits has been thoroughly investigated in \cite{Rizwan:2018rgs} for the case of rotating non-charged black hole in PFDM. Here, we provide an extensive study on the horizon when the black hole's charge $Q$ is also present. Without losing generality in methodology, we exclusively work with the positive values of $\lambda$ throughout the paper. Differing from the restriction of \(M^2 \ge a^2 + Q^2\) for the regular Kerr-Newman black hole, the presence of a dark matter background counteracts the negative gravitational effect caused by the electric charge. Consequently, larger values of \(Q\) are permissible in this context, provided that \(\lambda\) is sufficiently large to allow the logarithmic term in \(\Delta\) to turn negative. While it is impossible to obtain analytical expressions for \(r_{h\pm}\) and \(r_{L\pm}\) due to the complexity introduced by the logarithmic term in \(\Delta\), we perform a series of numerical studies on Eqn~\ref{eq:Delta} and Eqn~\ref{eq:gtt} with various combinations of $(a, Q, \lambda)$.

%\begin{widetext}
\begin{figure*}[htbp]
  \begin{center}
    \includegraphics[width=6.5in]{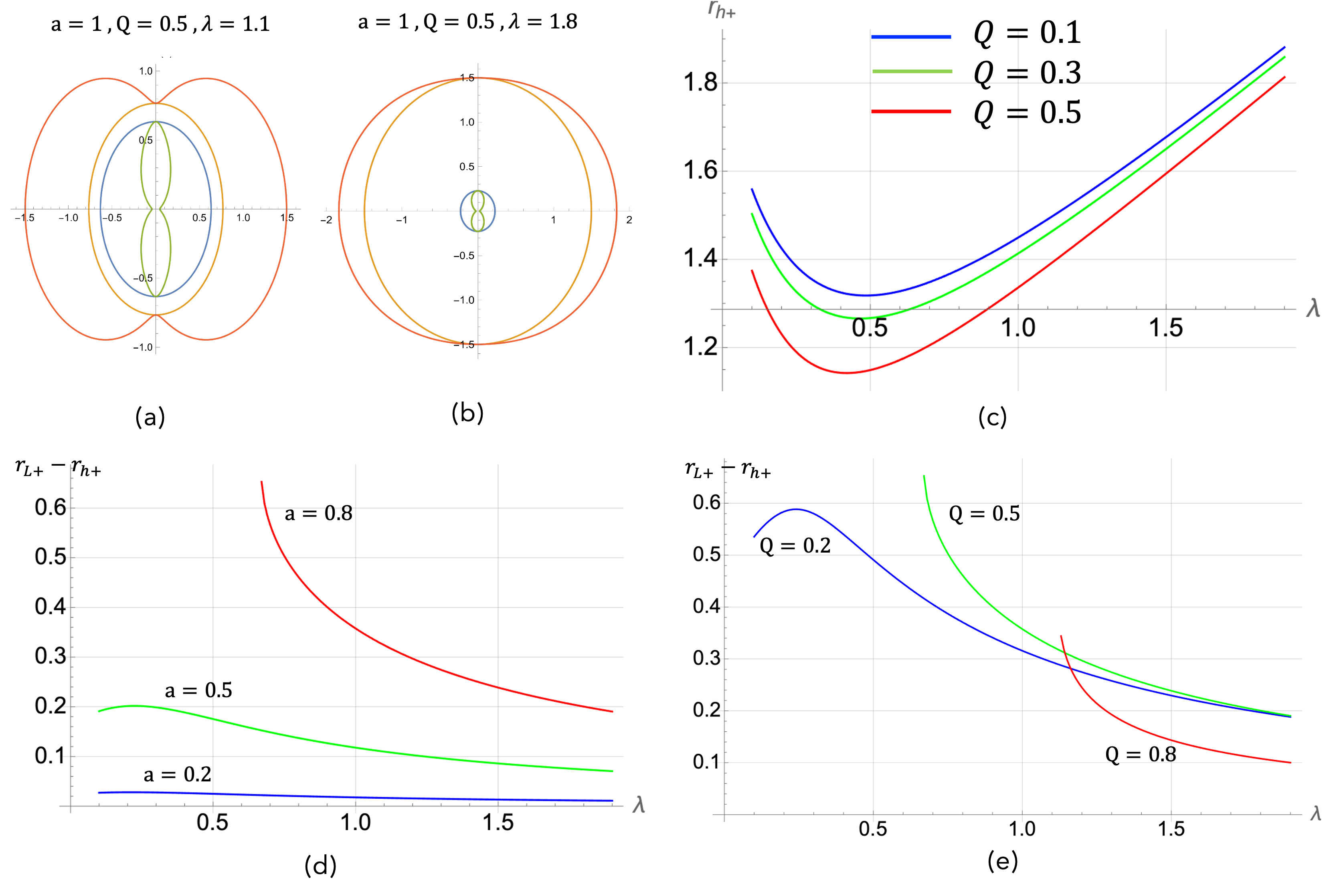}
 \caption{\label{fig:ergo} Panels (a) and (b) illustrate the event horizons and ergoregion for two values of dark matter parameter $\lambda=1.1$ and $1.8$ with fixed spin $a=1$ and electric charge $Q=0.5$. Red, orange, blue and green lines indicate the outer static limit $r_{L+}$, outer horizon $r_{h+}$, inner horizon $r_{h-}$ and inner static limit $r_{L-}$, respectively. Panel (c) shows the outer horizon as a function of the dark matter parameter for different charge values with spin $a=0.5$. Panels (d) and (e) show the size of ergoregion as a function of the dark matter parameter for different values of spin and charge. We set $Q=0.5$ in panel (d) and $a=0.8$ in panel (e).}
 \end{center}
\end{figure*}
%\end{widetext}

As illustrated in Fig~\ref{fig:ergo}, we observe several trends of how $(a, Q, \lambda)$ could affect the positions of the two horizons of this black hole, leading to significant changes in the size of the ergoregion. Panels (a) and (b) in Fig.~\ref{fig:ergo} show an example of a standard peanut-shaped ergoregion, bounded by the orange (outer horizon) and red (outer boundary of ergosphere) contours for two $\lambda$ values with $a$ and $Q$ fixed. A greater dark matter component could significantly increase the outer horizon, resulting in a shrinking ergoregion.  

Panel \ref{fig:ergo}(c) shows the outer horizon $r_{h+}$ as a function of the dark matter parameter $\lambda$ for three values of electric charge when the spin is fixed at $a=0.5$. Interestingly, the outer horizon initially decreases as $\lambda$ increases, then begins to rapidly increase with $\lambda$ once $\lambda$ surpasses a critical value $\lambda_c$ located at the minimum of $r_{h+}$. This critical value $\lambda_c$, interpreted as the point of reflection, has also been noted in previous studies \cite{Das:2020yxw} and has been observed for the shadow of the rotating black hole in PFDM \cite{Hou:2018avu}.

In panels \ref{fig:ergo}(d) and \ref{fig:ergo}(e), we show the size of the ergoregion, given as $r_{L+}-r_{h+}$, as a function of dark matter parameter $\lambda$ for varying spin and charge values. The variation of $r_{L+}-r_{h+}$ shows a similar trend to that of the outer horizon for lower to intermediate values of spin and charge. Interestingly, the size of ergoregion can be significantly elevated at faster spins ($a>0.8$) as the dark matter parameter $\lambda$ decreases, provided the electric charge stays at lower values ($Q<0.5$). These results provide us with a wide window of $(a, Q,\lambda)$ parameters to explore in the subsequent study of energy extraction in the ergosphere of the black hole under consideration. These observed influences from the dark matter component and electric charge are logical because both the dark matter dominance ($\lambda>\lambda_c$) and the presence of charge $Q$ can reduce the effective mass of the black hole system, leading to the effective relative enlargement of the outer horizon. 

Magnetic reconnection occurs within the bulk plasma orbiting circularly around the black hole, with circular orbits extending from infinity down to the limiting circular photon orbit. We now turn our attention to examine how the electric charge and dark matter background impact the circular null geodesics of the Kerr-Newman black hole in PFDM. We start by considering a test particle with Lagrangian
\begin{gather}
\mathcal{L}=\frac{1}{2}g_{\mu\nu}\dot{x}^{\mu}\dot{x}^{\nu},
\end{gather}
where the derivative is with respect to the proper time $\tau$. Since $\mathcal{L}$ is independent of $t$ and $\phi$, both the energy per unit mass $E$ and z-component angular momentum per unit mass $l_z$ are conserved for an observer at infinity. We can then represent the equations of motion for $t$ and $\phi$ as
\begin{align}
\begin{split}
\dot{t}&=\frac{g_{\phi\phi}E+g_{t\phi}l_z}{g_{t\phi}^2-g_{tt}g_{\phi\phi}}\\&=\frac{1}{r^2}\left [\frac{r^2+a^2}{\Delta}\left (E(r^2+a^2)-al_z\right )+a(l_z-aE)\right],\\
\dot{\phi}&=-\frac{g_{t\phi}E+g_{tt}l_z}{g_{t\phi}^2-g_{tt}g_{\phi\phi}}\\&=\frac{1}{r^2}\left [\frac{a}{\Delta}\left (E(r^2+a^2)-al_z\right )+(l_z-aE)\right].
\end{split}
\end{align}
The radial equation of motion can then be obtained from the conservation of the Hamiltonian $\mathcal{H}=p_{\mu}\dot{x}^{\mu}-\mathcal{L}$, yielding
\begin{gather}
\dot{r}^2=\frac{1}{r^4}\left [\left(E(r^2+a^2)-al_z\right )^2-\Delta (l_z-aE)^2\right ]+\frac{\Delta}{r^2} \epsilon,
\end{gather}
where $\epsilon=-1,0,1$ for timelike, null and spacelike particles, respectively. In the case of null geodesics, the radial equation becomes
\begin{align}
\begin{split}
\dot{r}^2=\frac{1}{r^4}\left [\left(E(r^2+a^2)-al_z\right )^2-\Delta (l_z-aE)^2\right ]=f(r)
\end{split}
\end{align}

\begin{figure*}[htbp]
  \begin{center}
    \includegraphics[width=6.5in]{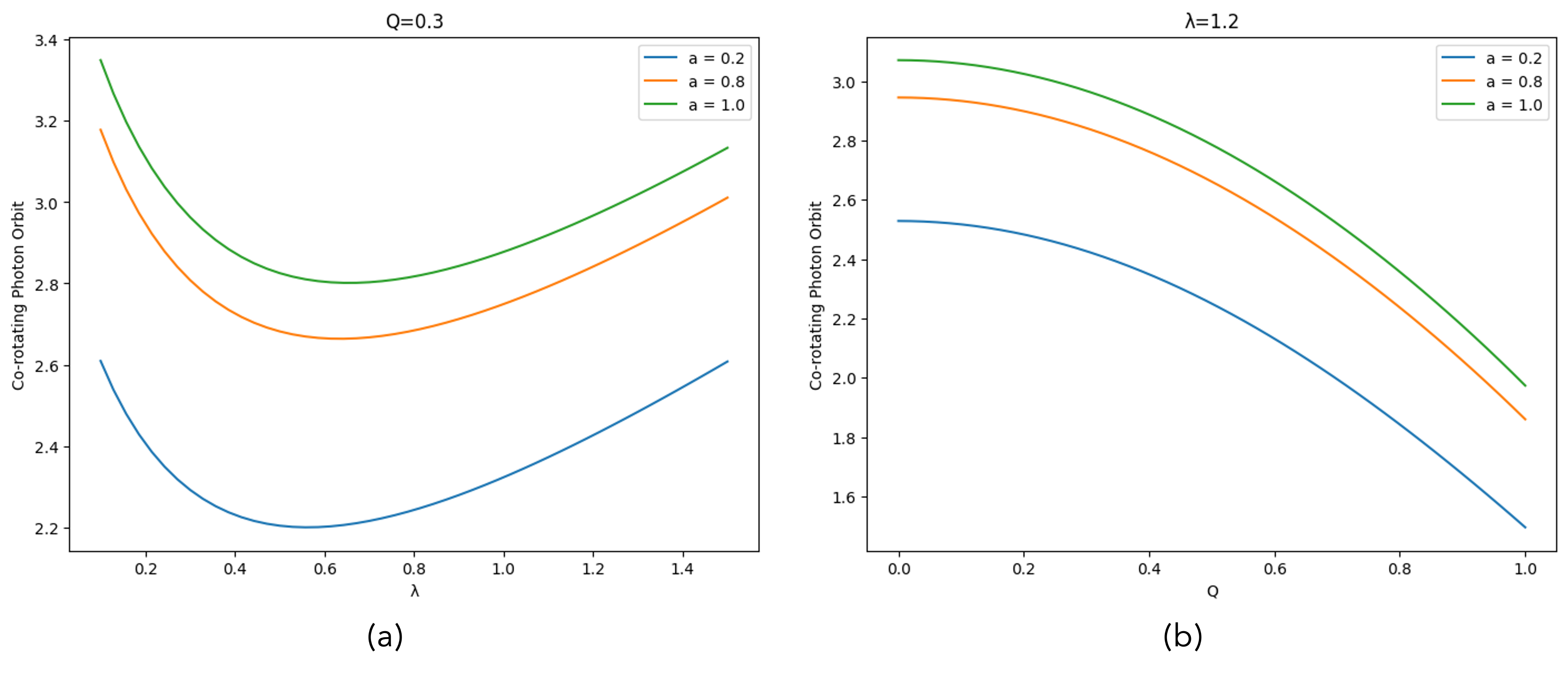}
 \caption{\label{fig:rp} Panel (a) illustrates the co-rotating photon orbit as a function of dark matter parameter $\lambda$ with fixed black hole charge $Q=0.3$ and three spin values $a=0.2, 0.8, 1.0$. Panel (b) illustrates the co-rotating photon orbit as a function of black hole charge $Q$ with fixed dark matter parameter $\lambda=1.2$ and three spin values $a=0.2, 0.8, 1.0$.}
 \end{center}
\end{figure*}

For circular orbits, we require$f(r)=f'(r)=0$. Let $D=l_z/E$ be the impact parameter, these two conditions can be rewritten as \cite{Das:2020yxw} 
\begin{align}
\begin{split}
r_p^2+\frac{2M}{r_p}(a-D)^2-\frac{Q^2}{r_p^2}(a-D)^2-\frac{\lambda}{r_p} \ln\left (\frac{r_p}{|\lambda|}\right ) (a-D)^2+(a^2-D^2)&=0,\\
2r_p-\frac{2M}{r_p^2}(a-D)^2+\frac{2Q^2}{r_p^3}(a-D)^2+\frac{\lambda}{r_p^2} \ln\left (\frac{r_p}{|\lambda|}\right ) (a-D)^2-\frac{\lambda}{r_p^2}(a-D)^2&=0.
\end{split}
\end{align}
Solving for $D$ and subtituting it into the first equation above yields the following equation for the co-rotating photon orbit $r_p$ (here we focus on co-rotating orbit for magnetic reconnection occurring inside the ergosphere)
\begin{align}
\begin{split}
6M r_p^3-4Q^2r_p^3-3\lambda r_p^3 \ln\left (\frac{r_p}{|\lambda|}\right ) +\lambda r_p^3-2r_p^4+2a\sqrt{2r_p^4\left ( 2Mr_p-2Q^2-\lambda r_p  \ln\left (\frac{r_p}{|\lambda|}\right )+\lambda r_p \right )} =0
\end{split}
\end{align}

Since obtaining an analytical expression for the $r_p$ is not feasible, we report numerical results for the co-rotating photon orbit. As shown in panel (a) of Fig~\ref{fig:rp}, the photon orbit $r_p$ exhibits a similar trend to the outer horizon in its dependence on the dark matter parameter $\lambda$ while keeping charge $Q$ fixed- it initially decreases with increasing $\lambda$ until reaching a critical value $\lambda_c$. Meanwhile, panel (b) of Fig~\ref{fig:rp} shows that $r_p$ decreases monotonically with increasing value of charge $Q$ while keeping dark matter parameter $\lambda$ fixed. These trends also align with the fact that when dark matter interacts with black holes, it effectively increases the mass, acting as an attractive gravitational charge (when $\lambda>\lambda_c$) that mitigates the repulsive effect of the electric charge. The results from this section clearly demonstrate the strong dependence of the Kerr-Newman black hole in PFDM spacetime on the black hole charge $Q$ and the dark matter parameter $\lambda$. Therefore, it is compelling to investigate their influences on the rate and efficiency of energy extraction via magnetic reconnection, which we explore in the next section.

\section{ENERGY EXTRACTION BY CAMISSO-ASENJO PROCESS}
To investigate the mechanism behind the energetic bursts of flares caused by charged particles near a black hole, we examine the energy extraction process via magnetic reconnection, as recently proposed by Comisso and Asenjo \cite{Comisso:2020ykg}. Specifically, we focus on the strong frame-dragging effect in a rotating charged black hole immersed in PFDM and analyze the efficiency of its rotational energy being extracted by the surrounding plasma in the accretion disk. First, we consider a locally nonrotating frame known as the Zero Angular Momentum Observer (ZAMO) frame \cite{Bardeen:1972fi}, which we will denote by hats for the rest of the paper. The spacetime in the ZAMO frame is locally flat (Minkowski) and its squared line element is given by
\begin{gather}
ds^2=-d\hat{t}^2+\Sigma^3_{i=1}(d\hat{x}^i)^2=\eta_{\mu\nu}d\hat{x}^{\mu}d\hat{x}^{\nu},
\end{gather}
The coordinate transformation between the ZAMO frame ($\hat{t},\hat{x}^i$) and the Boyer-Lindquist coordinates ($t,x^i$) is $d\hat{t}=\alpha dt$ and $d\hat{x}^i=\sqrt{g_{ii}}dx^i-\alpha\beta^i dt$, where the lapse function is given by $\alpha=\sqrt{-g_{tt}+\frac{g_{t\phi}^2}{g_{\phi\phi}}}$ and the shift vector is given by $\beta^i=(0,0,\frac{\sqrt{g_{\phi\phi}}\omega_{\phi}}{\alpha})$, $\omega_{\phi}=-g_{t\phi}/g_{\phi\phi}$ being the angular velocity of the frame dragging.

Considering plasma fluid co-rotates around the equatorial plane of this rotating charged black hole in PFDM, the Keplarian velocity in Boyer-Lindquist coordinates can be obtained from

\begin{align}
\begin{split}
\Omega_K =\frac{d\phi}{dt}=\frac{-\partial_r g_{t\phi}\pm\sqrt{(\partial_r g_{t\phi})^2-(\partial_r g_{tt})(\partial_r g_{\phi\phi})}}{\partial_r g_{\phi\phi}}
\end{split}
\end{align}

Note that contravariant vectors in the ZAMO frames can be transformed to the Boyer-Lindquist coordinates by $\hat{b}^0=\alpha b^0$, and $\hat{b}^i=\sqrt{g_{ii}}b^i-\alpha \beta^ib^0$, while covectors transform by $\hat{b}_0=b_0/\alpha+\Sigma^3_{i=1}(\beta^i/\sqrt{g_{ii}})b_i$ and $\hat{b}_i=b_i/\sqrt{g_{ii}}$. With these transformations, the Keplarian velocity of the corotating bulk plasma in ZAMO frame can be found by $\hat{v}_K=\frac{d\hat{x}^{\phi}}{d\hat{t}}=\frac{\sqrt{g_{\phi\phi}}}{\alpha}{\Omega_K}-\beta^{\phi}$.

Assuming plasma properties in the magnetic reconnection process are described using the energy-momentum tensor in the one-fluid approximation
\begin{gather}
T^{\mu\nu}_b=pg^{\mu\nu}+\omega U^{\mu}U^{\nu}+F\indices{^{\mu}_{\delta}} F^{\nu\delta}-\frac{1}{4}g^{\mu\nu}F^{\rho\delta}F_{\rho\delta},
\end{gather}
where $\rho$, $\omega$, $U^{\mu}$, and $F^{\mu\nu}$ are the proper plasma pressure, enthalpy density, four-velocity, and the electromagnetic field tensor, respectively. Note that we use $T^{\mu\nu}_b$ here to distinguish it from the energy-momentum tensor $T^s_{\mu\nu}$ that sources the black hole. We assume the spacetime is fixed and the bulk plasma does not affect the black hole's charge $Q$. Therefore, the electromagnetic field tensor $F_{\mu\nu}$ in the above equation refers to the contribution from the bulk plasma only. 

Similar to the Penrose process, energy extraction via magnetic reconnection to occur at the ergosphere, the accelerated plasma escaping the black hole attains positive energy, while the decelerated plasma absorbed by the black hole must hold negative energy. To evaluate the energy measured by static observers at infinity, we construct a quantity, known as the ``energy-at-infinity" density $e^{\infty}$
\begin{gather}
 e^{\infty}\equiv n_{\mu}J^{\mu}=n_{\mu}T^{\mu\nu}_b\xi_{\nu}=-\alpha g_{\mu 0}T^{\mu 0}_b=\alpha \hat{e}+\alpha \beta^\phi \hat{P}^\phi,  
 \end{gather}
where $\xi_\nu=(\partial_t, 0, 0, 0)$ is the time-like Killing vector and $n_{\nu}$ is the unit vector normal to time-like hypersurfaces. The total energy density $\hat{e}$ and azimuthal component of the momentum density $\hat{P}^{\phi}$ are given as
\begin{align}
\begin{split}
\hat{e}&=\omega \hat{\gamma}^2-p+\frac{\hat{B}^2+\hat{E}^2}{2},\\
\hat{P}^\phi&=\omega \hat{\gamma}^2\hat{v}^\phi + (\hat{B}\times\hat{E})^\phi,
\end{split}
\end{align}
where the outflow Lorentz factor $\hat{\gamma}=\hat{U}^0=\sqrt{1-\Sigma^3_{i=1}(d\hat{v}^i)^2}$ and $\hat{v}^{\phi}$ is the azimuthal component of the outflow velocity measured by the ZAMO observer. The electric and magnetic field components are defined by
\begin{align}
\begin{split}
\hat{B}^i&=\frac{1}{2}\epsilon^{ijk}F_{jk},\\
\hat{E}^i&=\eta^{ij}F_{j0}.
\end{split}
\end{align}
The energy density at infinity can be further rewritten in terms of its hydrodynamic and electromagnetic components: $e^\infty=e^\infty_{hyd}+e^\infty_{em}$, expressed as
\begin{align}
\begin{split}
e^\infty_{hyd}&=\alpha \hat{e}_{hyd}+\alpha \beta^\phi \omega \hat{\gamma}^2\hat{v}^\phi,\\
e^\infty_{em}&=\alpha \hat{e}_{em}+\alpha\beta^\phi (\hat{B}\times\hat{E})_\phi,
\end{split}
\end{align}
where $\hat{e}_{hyd}=\omega \hat{\gamma}^2-p$ and $\hat{e}_{em}=(\hat{B}^2+\hat{E}^2)/2$ are the energy densities of the hydrodynamic and electromagnetic fields in the ZAMO frame. Adopting Comisso and Asenjo's assumption \cite{Comisso:2020ykg}, we consider the bulk plasma is incompressible and adiabatic, and the conversion of magnetic energy into plasma's kinetic energy during magnetic reconnection is highly efficient. Under this condition, the energy density at infinity $e^{\infty}$ is predominantly dominated by the hydrodynamic component, with the electromagnetic contribution being negligible, leading to an expression
\begin{gather}
%e^{\infty}=\alpha \hat{e}+\alpha \beta^{\phi}\hat{P}^{\phi},
e^{\infty}\approx \alpha\hat{\gamma}\omega (1+\beta^\phi\hat{v}^{\phi})-\frac{\alpha p}{\hat{\gamma}}, \label{eq:einfty}
\end{gather}

 \begin{figure}[h!] 
  \begin{center}
    \includegraphics[width=6.5in]{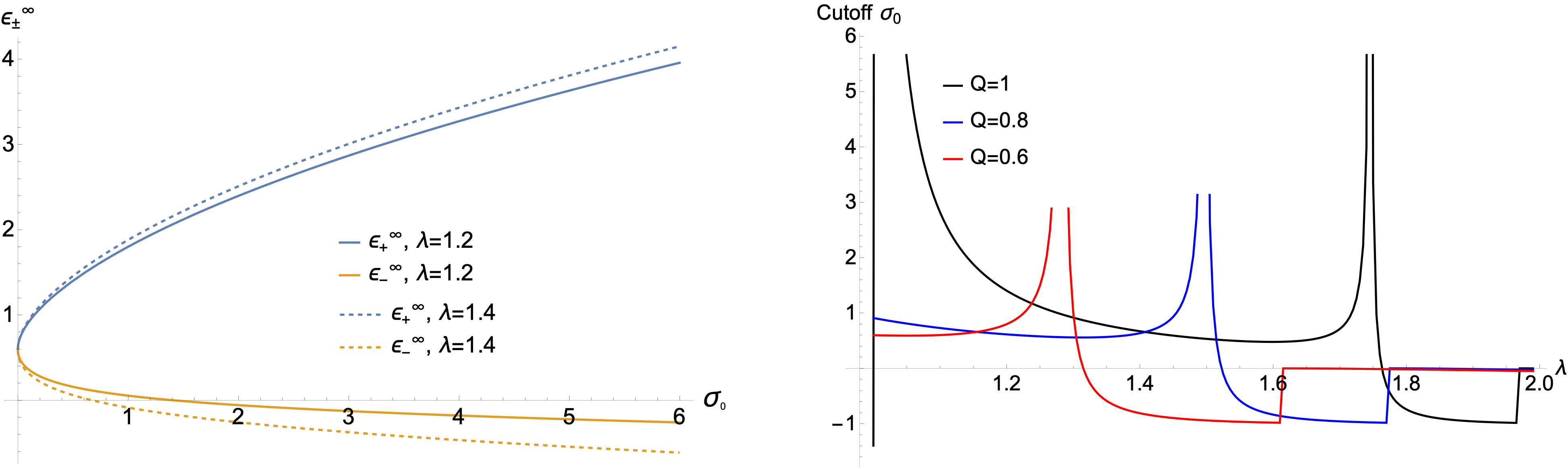}
    %\captionsetup{justification=justified, singlelinecheck=false}
    \caption{\label{fig:sigma2} Left figure: energy-at-infinity density per enthalpy, $\epsilon^\infty_{+}$ (blue lines) and $\epsilon^\infty_{-}$ (orange lines) as a function of plasma magnetization $\sigma_0$, for two $\lambda$ values with $(a, Q, \xi)$ fixed at $(1.0, 1.0, 0)$. Right figure: cutoff values of $\sigma_0$ as a function of $\lambda$ for three $Q$ values.}
 \end{center}
 \end{figure}

In Comisso and Asenjo's configuration, the current density is directed radially within the reconnection layer in the azimuthal direction. The outflow velocity and its associated Lorentz factor are expressed in terms of the plasma magnetization upstream of the reconnection layer $\sigma_0$
\begin{gather}
v_{out}\approx\left (\frac{\sigma_0}{1+\sigma_0}\right )^{1/2}, \label{eq:vout}\\
~~\gamma_{out}=(1-v_{out}^2)^{-1/2}\approx(1+\sigma_0)^{1/2}, \label{eq:gammaout}
\end{gather}
where $\sigma_0=B_0^2/\omega_0$, and $B_0$ and $\omega_0$ are the asymptotic macro-scale magnetic field and plasma enthalpy density, respectively. The azimuthal components of the two outflow velocities are then expressed as
\begin{gather}
v^\phi_{\pm}=\frac{\hat{v}_K\pm v_{out} \cos \xi}{1\pm \hat{v}_K v_{out}\cos\xi}, \label{eq:vphi}
\end{gather}
where $\pm$ represents the corotating $(+)$ and counter-rotating $(-)$ plasma flow relative to the black hole, and $\xi$ denotes the angle between the radial and azimuthal components of the outflow velocity. Combining equations~\ref{eq:einfty} to \ref{eq:vphi}, the final expression for the energy density at infinity per enthalpy $\epsilon^\infty=e^\infty/\omega$ leads to \cite{Comisso:2020ykg}

\begin{align}
\begin{split}
\epsilon_{\pm}^{\infty}=\alpha\hat{\gamma}_K\biggl [ (1+\beta^{\phi}\hat{v}_K)(1+\sigma_0)^{1/2} \pm \cos\xi(\beta^{\phi}+\hat{v}_K)\sigma_0^{1/2} \\
%\begin{center}
-\frac{1}{4}\frac{(1+\sigma_0)^{1/2}\mp\cos\xi~\hat{v}_K\sigma_0^{1/2}}{{\hat{\gamma}_K}^2(1+\sigma_0-{\cos^2\xi}~{\hat{v}_K}^2\sigma_0)}\biggr ]
%\end{center}
\end{split}
\label{eq:eps}
\end{align}
Similarly, as in the Penrose process, energy extraction for the rotating charged black hole in PFDM occurs when the decelerated plasma secures negative energy-at-infinity, while the accelerated plasma particles possess energy-at-infinity more than its rest mass and thermal energies. This requires the following conditions to be met \cite{Comisso:2020ykg}
\begin{gather}
\epsilon^{\infty}_{-}<0,\\
\Delta \epsilon^{\infty}_{+}=\epsilon^{\infty}_{+}-\biggl (1-\frac{\Gamma}{\Gamma-1}\frac{p}{\omega}\biggr )=\epsilon^{\infty}_{+}>0
\end{gather}
where the polytropic index $\Gamma=4/3$ is assumed for relativistic hot plasma.

 \begin{figure}[h!] %Fig 
  \begin{center}
    \includegraphics[width=6.5in]{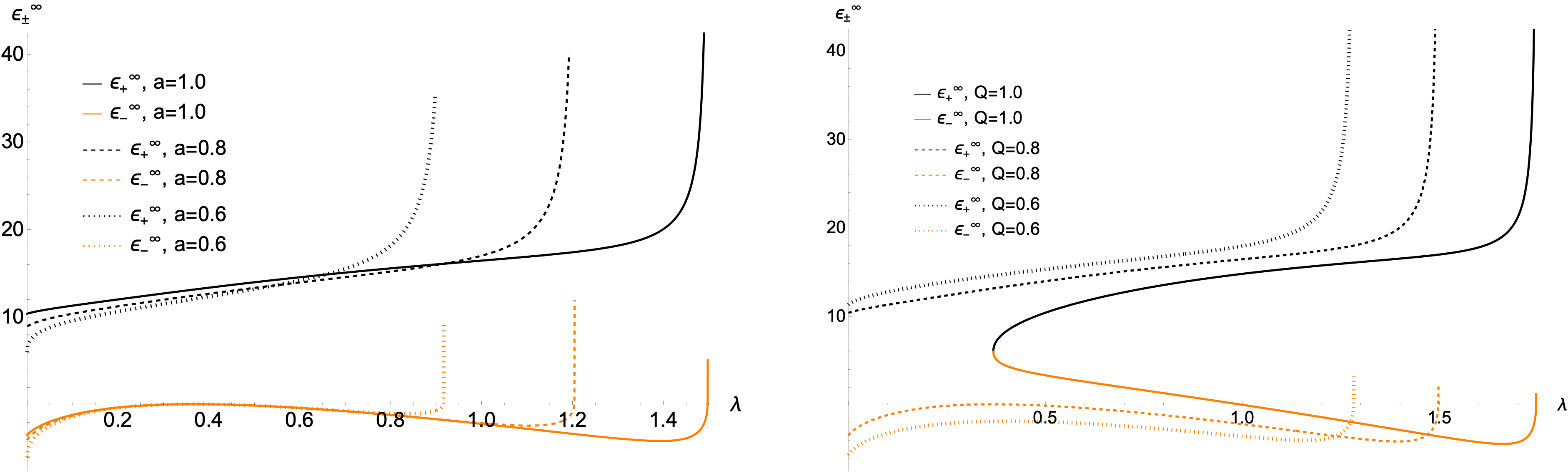}
 \caption{\label{fig:lambda2} Energy-at-infinity density per enthalpy, $\epsilon^\infty_{+}$ (black lines) and $\epsilon^\infty_{-}$ (orange lines), as a function of the dark matter parameter $\lambda$ for varying $a$ with fixed $Q$ at 0.8 (left figure), and varying $Q$ with fixed $a$ at 0.8 (right figure). For both cases, we choose $\xi=0$ and $\sigma_0=100$.}
 \end{center}
\end{figure}

\subsection{PARAMETER AND PHASE SPACE ANALYSIS}

To analyze the energy extraction by the CA process, we consider the above expression for the energy-at-infinity densities $\epsilon^\infty_{\pm}$, which are functions of the X-point location, $r/M$, and the parameter space $(a, Q, \lambda, \xi, \sigma_0)$. Since the energy extraction from various types of rotating black holes without charge or dark matter background has been studied in detail, our investigation focuses on how the PFDM parameter $\lambda$ and electric charge $Q$ affect the energy-at-infinity densities. 

As we consider various combinations of the parameters ($a, \xi, \sigma_0$), it is important to note that energy extraction is favored by lower values of the orientation angle $\xi$ and higher values of $\sigma_0$, as implied by equation \ref{eq:eps} and demonstrated in the standard Kerr case \cite{Comisso:2020ykg}. This condition is illustrated in the left figure of Fig.~\ref{fig:sigma2}, where $\epsilon^\infty_{+}$ (for accelerating plasma) and $\epsilon^\infty_{-}$ (for decelerating plasma) are shown for two values of the dark matter parameter $\lambda$. In both cases, $\epsilon^\infty_{-}$ becomes more negative as $\sigma_0$ increases, but the cutoff value of $\sigma_0$, beyond which $\epsilon^\infty_{-} < 0$ is secured, appears to decrease as $\lambda$ increases. The right figure of Fig.~\ref{fig:sigma2} shows that the dependence of the cutoff $\sigma_0$ on $\lambda$ is not always monotonic. Additionally, it indicates that $\lambda$ must remain below a certain critical value for a positive cutoff $\sigma_0$ to be possible. This is not surprising because, as previously discussed, such critical values for $\lambda$ can effectively influence horizons, which in turn determine the location of the reconnection X-point to occur within the ergoregion. 

\begin{figure}[h!] %Fig 
  \begin{center}
  \vspace{0.2cm}
    \includegraphics[width=3in]{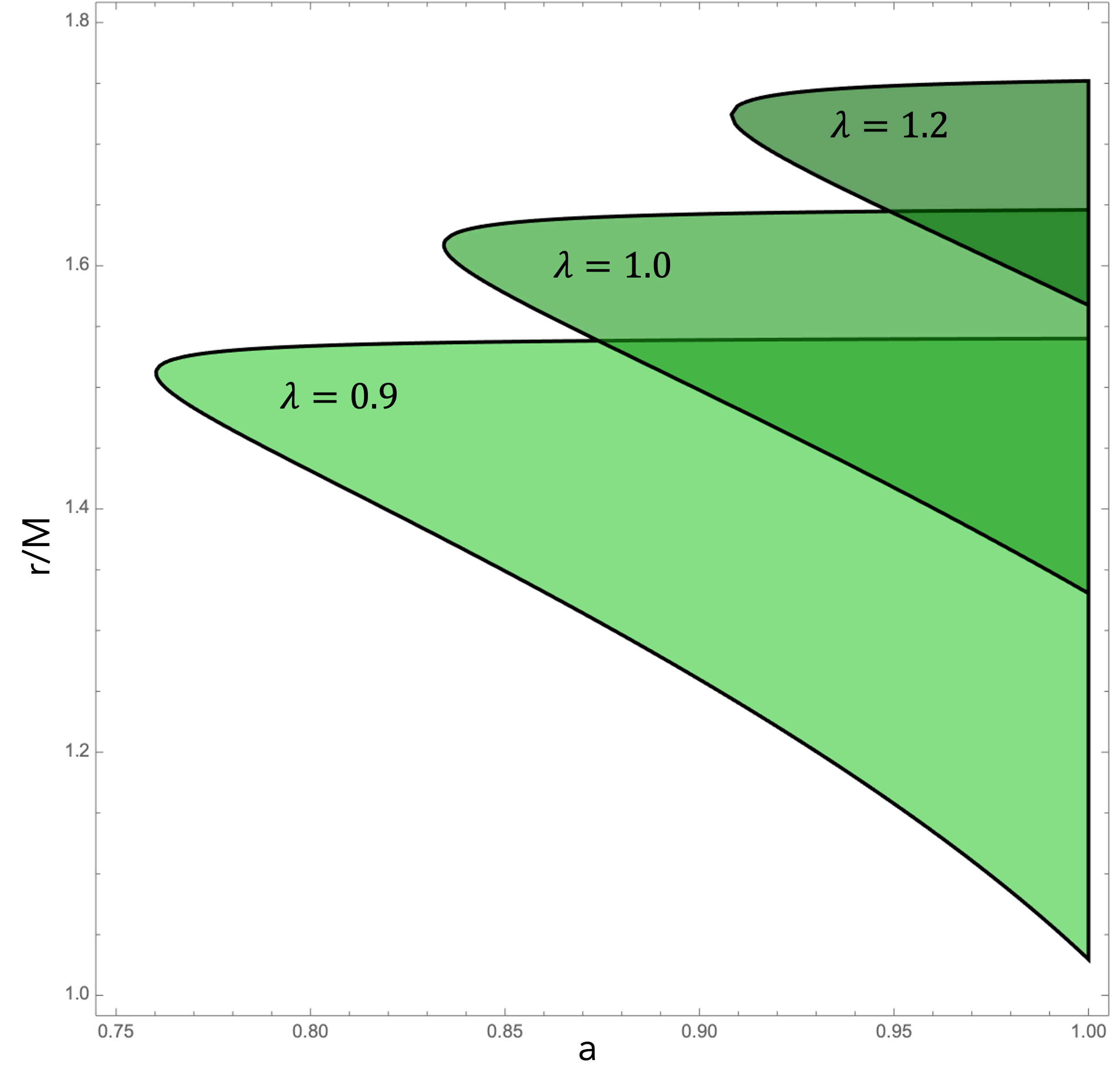}
 \caption{\label{fig:xpoint} Regions of phase-space $(a, r/M)$ where energy extraction occurs ($\epsilon^{\infty}_{-}<0$) for different values of the dark matter parameter $\lambda\in \{0.9, 1.0, 1.2\}$ with $(Q, \xi, \sigma)=(0.1, 0, 100)$. The areas with negative $\epsilon^{\infty}_{-}$ increase as $\lambda$ decreases, depicted in progressively lighter shades of green.}
 \end{center}
\end{figure}

The impact of dark matter on $\epsilon^\infty_{\pm}$ is further depicted in Fig~\ref{fig:lambda2} for different combinations of $a$ and $Q$. Both the left and right figures indicate that $\epsilon^\infty_{-}<0$ is met within a range of $\lambda$ values that widens as $Q$ decreases and $a$ increases. The values of $\lambda$ in these ranges increase monotonically as either $a$ or $Q$ increases. Fig~\ref{fig:xpoint} illustrates the strong influence of the dark matter parameter on the regions of phase-space $(a,r/M)$ where energy extraction occurs $(\epsilon^\infty_{-}<0)$. It shows a clear trend that the phase-space region shifts to a larger radial distance as $\lambda$ increases and this region also expands as $\lambda$ decreases. The key point to take away is that the phase-space region can extend to a larger radial location where energy extraction favors due to the combined effect of the black hole charge and dark matter parameter.

\subsection{ENERGY EXTRACTION RATE AND RECONNECTION EFFICIENCY}
In this section, we continue to analyze the role of the parameter space on the energy extraction rate $P_{extr}$ and reconnection efficiency $\eta$. The energy extraction rate is determined by the amount of plasma with negative energy at infinity falling into the ergosphere and can be estimated as \cite{Comisso:2020ykg}
\begin{align}
\begin{split}
P_{extr}=-\epsilon^{\infty}_{-}\omega_0 A_{in} U_{i}.
\end{split}
\end{align}
Here $A_{in}\approx r^2_{L+}-r^2_{isco} $ is the cross-section area of the inflowing plasma, and we use $U_{in}\approx 0.1$ for magnetic reconnection in the collisionless regime. In the CA mechanism, reconnection is responsible for generating fast plasma outflows that extract energy from the black hole, so a highly efficient reconnection process can potentially induce a high energy extraction rate. The reconnection efficiency of the plasma energization process is readily defined as 
\begin{align}
\begin{split}
\eta=\frac{\epsilon^{\infty_{+}}}{\epsilon^{\infty}_{+}+\epsilon^{\infty}_{-}}.
\end{split}
\end{align}
The condition for energy extraction from the black hole is $\eta>1$. 

\begin{figure}[h!] %Fig
  \begin{center}
    \includegraphics[width=6.5in]{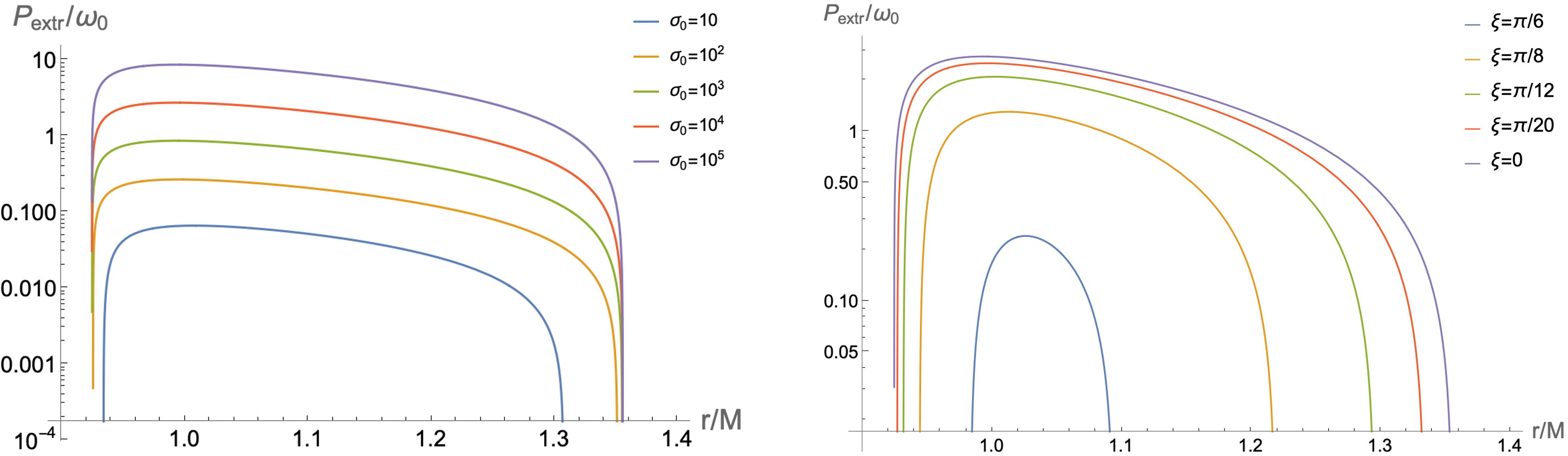}
 \caption{\label{fig:pext0} Extracted power per enthalpy $P_{extr}/\omega_0$ as a function of the X-point location with varying plasma magnetization (left figure) and varying orientation angle (right figure) with $(a, Q, \lambda)$ fixed at $(0.8, 0.8, 1.17)$.}
 \end{center}
\end{figure}

We first show in Fig.~\ref{fig:pext0} the extracted power per enthalpy, $P_{extr}/\omega_0$, as a function of the X-point location $r/M$ for different plasma magnetization values $\sigma_0\in \{10,10^2, 10^3, 10^4, 10^5\}$ with $(a, Q, \lambda, \xi)$ = $(0.8, 0.8, 1.17, 0)$ in the left figure, and for different orientation angle values $\xi \in \{0, \pi/20, \pi/12, \pi/8, \pi/6\}$ with $(a, Q, \lambda, \sigma_0)$ = $(0.8, 0.8, 1.17, 100)$ in the right figure. Similar to the standard Kerr case \cite{Comisso:2020ykg}, the power extracted from the black hole favors high plasma magnetization and low orientation angle values until it drops off near the limiting circular orbits. For the Kerr-Newman black hole in the presence of PFDM, we also observe the increased sensitivity of the orientation angle on the extracted power and its drop-off limits. For the remainder of the analysis, we use the preferred values of $\xi=0$ and $\sigma_0=10^4$ based on these results.

\begin{figure*}[htbp]
  \begin{center}
    \includegraphics[width=6in]{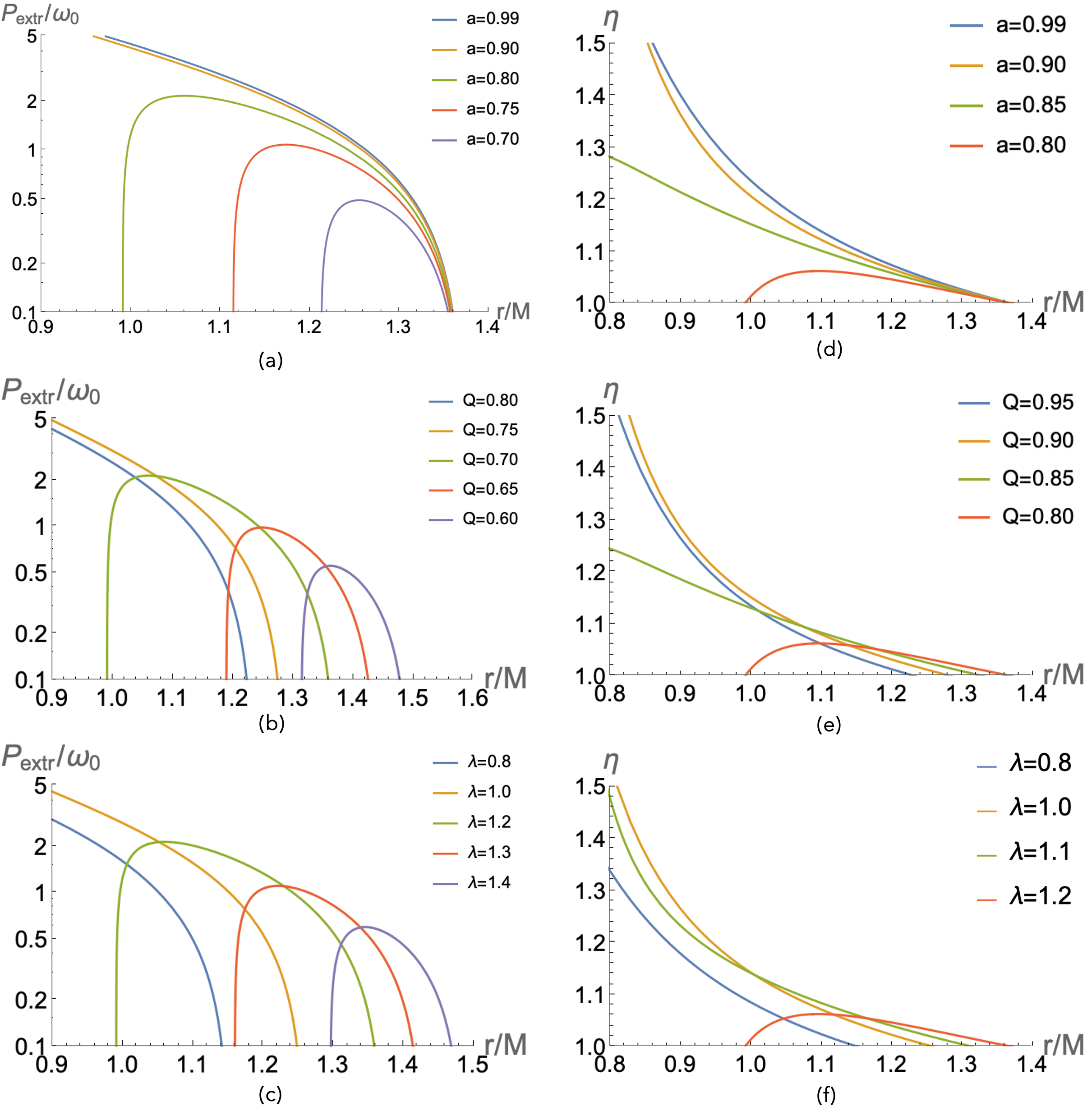}
 \caption{\label{fig:pexteta} The left three panels show the extracted power per enthalpy $P_{extr}/\omega_0$ as a function of the X-point location with (a): varying spin values but fixed $Q=0.8$ and $\lambda=1.2$; (b): varying electric charge values but fixed $a=0.8$ and $\lambda=1.2$; (c): varying dark matter parameters but fixed $a=0.8$ and $Q=0.8$. All with fixed $\xi=0$ and $\sigma_0=10^4$. Note that the green line practically represents the variation of  $P_{extr}/\omega_0$ with $(a, Q, \lambda)$ set at $(0.8, 0.8, 1.2)$ for panels (a)-(c). The right three panels show the reconnection efficiency $\eta$ as a function of the X-point location with (d): varying spin values but fixed $Q=0.8$ and $\lambda=1.2$; (e): varying electric charge values but fixed $a=0.8$ and $\lambda=1.2$; (f): varying dark matter parameters but fixed $a=0.8$ and $Q=0.8$. All panels are with fixed $\xi=0$ and $\sigma_0=10^4$. Note that the red line practically represents the variation of $\eta$ with $(a, Q, \lambda)$ set at $(0.8, 0.8, 1.2)$ for panels (d)-(f).}
 \end{center}
\end{figure*}

In Fig.~\ref{fig:pexteta}(a)-(c), we show the ratio $P_{extr}/\omega_0$ as a function of the X-point location for varying values of spin, electric charge, and dark matter parameter. The variation trends of $P_{extr}/\omega_0$ can be categorized into two types. For lower black hole spins, lower electric charges, and higher dark matter components, the extracted power increases with increasing black hole spin and electric charge, but decreases with a higher dark matter parameter until it drops off rapidly at the limiting circular orbits. Conversely, this trend becomes monotonic when the black hole spin approaches unity, with higher charges or greater dark matter dominance. To conveniently compare the values of $P_{extr}/\omega_0$, we use the green line as a reference when the parameters $(a, Q, \lambda)$ are held at $(0.8, 0.8, 1.2)$ for panels Fig.~\ref{fig:pexteta}(a)-(c). Interestingly, as shown in Fig.~\ref{fig:pexteta}(b) and (c) with $a=0.8$ fixed, although the extracted power is not favored at lower values of black hole spin, by either increasing either the electric charge or decreasing the dark matter contribution, $P_{extr}/\omega_0$ can be elevated to the same high level of near-unity spin cases. 

In Fig.~\ref{fig:pexteta}(d)-(f), the reconnection efficiency $\eta$ is depicted as a function of the X-point location $r/M$ for different values of spin, electric charge and dark matter parameter. The efficiency follows similar trends to those observed for the extracted power: $\eta$ rapidly approaches $\eta_{\approx}1.5$ (comparable to the regular Kerr case) near the outer horizon for higher values of spin and charge, and lower values of dark matter component. Fig.~\ref{fig:pexteta}(d) shows that, as the spin decreases, the reconnection efficiency significantly falls below unity. However, Fig.~\ref{fig:pexteta}(e) and (f) show that, with the spin fixed at a relatively low value of 0.8, the reconnection efficiency can be elevated back up to $\eta_{max}$ by either turning up the electric charge or turning down the dark matter contribution. 

The fact that parameters $(a, Q,\lambda)$ play the same role in determining both  $P_{extr}/\omega_0$ and $\eta$ confirms the close connection between energy extraction from the black hole and the magnetic reconnection process. More importantly, this new feature, arising particularly in the Kerr-Newman black hole in the presence of PFDM, allows magnetic reconnection to serve as an energy extraction mechanism for less rapidly rotating black holes (e.g. $a=0.8$), as effectively as in near-unity spin cases by incorporating electric charge and the dark matter component.

 %-----------------------------
\section{Summary and Conclusions}
As proposed by Comisso and Ansenjo \cite{Comisso:2020ykg}, magnetic reconnection can serve as an efficient mechanism for energy extraction from a rotating black hole in the ergosphere, where frame dragging causes anti-parallel magnetic field lines and a current sheet to form. Magnetic reconnection converts magnetic energy into plasma particle energy, producing rapid outflows; energy extraction occurs when the outflow moving opposite to the black hole's rotation has negative energy-at-infinity, while the other outflow escapes to infinity with gained energy. This novel way of extracting a black hole's rotational energy via rapid energy release during reconfiguration of the magnetic field has provided new insights into explaining bursty relativistic jets.

In this work, we evaluated this process for a Kerr-Newman black hole in the presence of perfect fluid dark matter, whose spacetime involves two new parameters, electric charge $Q$ and the dark matter density parameter $\lambda$, in addition to the standard parameter spin $a$. Our studies show that the combination of $(a,Q, \lambda)$ can significantly impact the horizons, the size of ergoregion, and circular geodesics of the black hole, leading to their important influences in energy extraction conditions, energy extraction rate, and reconnection efficiency. 

We first studied the geodesics of the black hole in question and looked for optimized cases where its ergoregion is reasonably sizable for energy extraction. Since the metric components $g_{tt}$ and $g_{rr}$ involve a logarithmic term, analyzing the event horizons and circular geodesics becomes much more complex and can only be performed numerically. As summarized in Fig~\ref{fig:ergo} and Fig~\ref{fig:rp}, we found that the event horizons and circular photo orbits strongly depend on both the electric charge $Q$ and the dark matter parameter $\lambda$. Particularly, the outer horizon $r_{h+}$ and co-rotating photon orbit $r_p$ initially decreases as $\lambda$ increases until reaching a critical value $\lambda_c$, beyond which $r_{h+}$ and $r_p$ increase more rapidly as $\lambda$ grows. A similar non-monotonic trend holds for the variation of the size of ergoregion $r_{L+}-r_{h+}$ as a function of $\lambda$. We also found that the size of ergoregion can significantly increase at faster spins ($a>0.8$) as the dark matter parameter $\lambda$ decreases, given the electric charge stays within $Q\in [0.2,0.5]$. 

Next, we evaluated the energy extraction conditions: energy-at-infinity per enthalpy for decelerating plasma $\epsilon^{\infty}_{-}<0$ and energy-at-infinity per enthalpy $\epsilon^{\infty}_{+}>0$ for various combinations of parameters involved $(a,Q,\lambda, \xi,\sigma_0)$. Similar to the standard Kerr case, energy extraction in Kerr-Newman black hole in PFDM is also favored at higher values of plasma magnetization $\sigma_0$ and near-null orientation angle $\xi$. However, the presence of charge $Q$ and dark matter component $\lambda$ introduces additional features in determining the cutoff value of plasma magnetization that triggers energy extraction. As depicted in Fig~\ref{fig:sigma2}, we found that this cutoff plasma magnetization can be lowered if the dark matter component increases or the charge decreases. Fig~\ref{fig:lambda2} shows that the permitted range of dark matter contribution for energy extraction to occur strongly depends on both the black hole spin and charge. Additionally, a lower dark matter component results in greater regions in phase-space $(a, r/M)$, while increasing the dark matter parameter can also extend the phase-space region to a larger radial distance favored by energy extraction.

Lastly, we analyzed the power extracted per enthalpy $P_{ext}/\omega_0$ and reconnection efficiency $\eta$ as a function of X-point location for different values in the full parameter space $(a, Q, \lambda, \xi,\sigma_0)$. Figs~\ref{fig:pext0}, \ref{fig:pexteta}(a) and \ref{fig:pexteta}(d) confirm that extracted power is still enhanced by high plasma magnetization, low orientation angle and high black hole spin, with the maximum value of $P_{ext}/\omega_0\approx 5$ and $\eta\approx 1.5$ achieved at near-unity spins comparable to the standard near-extremal Kerr case. Most interestingly, we found that the presence of charge and dark matter background could elevate $P_{ext}/\omega_0$ and $\eta$ to the same maximum values even when the black hole spin is not particularly high, as shown in panels (b)-(c) and (e)-(f) in Fig.\ref{fig:pexteta}. It relaxes the stringent requirements found in other types of rotating black holes \cite{Wang:2022qmg, Liu:2022qnr, Li:2023htz, Li:2023nmy, Zhang:2024rvk, Khodadi:2022dff, Khodadi:2023juk, Zhang:2024ptp, Shaymatov:2023dtt, Carleo:2022qlv}, where near-extremal spin is typically necessary to achieve a comparable level of extracted power and reconnection efficiency.

An interesting avenue to pursue next is to explore how energy extraction via magnetic reconnection transpires in accretion disks surrounding black holes in conformal Weyl gravity. It is well known that black holes in this theory naturally encode dark matter and dark energy within their vacuum, without requiring adding dark matter and dark energy terms in the Lagrangian \cite{Mannheim:1988dj}. Additionally, conformal Weyl gravity includes the Kerr spacetime as one of its vacuum solutions, allowing the CA framework to be naturally applied. Extending energy extraction studies to the more general conformal Weyl gravity, which enhances general relativity with an additional conformal symmetry, could reveal new and novel aspects of magnetic reconnection as an energy extraction mechanism from rotating black holes.

\section*{Acknowledgments}
SR would like to thank the Grinnell College Harris Fellowship Foundation for supporting this work and also extend gratitude to WPI for their hospitality during its completion. SR would also like to thank Luca Comisso for his support, encouragement, and very enlightening discussions at the inception of this work. AS, SR, and LR would like to thank Grinnell College CSFS funding for their support of this work.

\bibliographystyle{utphys} 
\bibliography{cftgr}

\providecommand{\href}[2]{#2}\begingroup\raggedright\begin{thebibliography}{10}

\bibitem{LIGOScientific:2016aoc}
{\bfseries LIGO Scientific, Virgo} Collaboration, B.~P. Abbott {\em et~al.},
  ``{Observation of Gravitational Waves from a Binary Black Hole Merger},''
  \href{http://dx.doi.org/10.1103/PhysRevLett.116.061102}{{\em Phys. Rev.
  Lett.} {\bfseries 116} no.~6, (2016) 061102},
  \href{http://arxiv.org/abs/1602.03837}{{\ttfamily arXiv:1602.03837 [gr-qc]}}.

\bibitem{LIGOScientific:2019fpa}
{\bfseries LIGO Scientific, Virgo} Collaboration, B.~P. Abbott {\em et~al.},
  ``{Tests of General Relativity with the Binary Black Hole Signals from the
  LIGO-Virgo Catalog GWTC-1},''
  \href{http://dx.doi.org/10.1103/PhysRevD.100.104036}{{\em Phys. Rev. D}
  {\bfseries 100} no.~10, (2019) 104036},
  \href{http://arxiv.org/abs/1903.04467}{{\ttfamily arXiv:1903.04467 [gr-qc]}}.

\bibitem{EventHorizonTelescope:2019ths}
{\bfseries Event Horizon Telescope} Collaboration, K.~Akiyama {\em et~al.},
  ``{First M87 Event Horizon Telescope Results. IV. Imaging the Central
  Supermassive Black Hole},''
  \href{http://dx.doi.org/10.3847/2041-8213/ab0e85}{{\em Astrophys. J. Lett.}
  {\bfseries 875} no.~1, (2019) L4},
  \href{http://arxiv.org/abs/1906.11241}{{\ttfamily arXiv:1906.11241
  [astro-ph.GA]}}.

\bibitem{EventHorizonTelescope:2019dse}
{\bfseries Event Horizon Telescope} Collaboration, K.~Akiyama {\em et~al.},
  ``{First M87 Event Horizon Telescope Results. I. The Shadow of the
  Supermassive Black Hole},''
  \href{http://dx.doi.org/10.3847/2041-8213/ab0ec7}{{\em Astrophys. J. Lett.}
  {\bfseries 875} (2019) L1}, \href{http://arxiv.org/abs/1906.11238}{{\ttfamily
  arXiv:1906.11238 [astro-ph.GA]}}.

\bibitem{EventHorizonTelescope:2022wkp}
{\bfseries Event Horizon Telescope} Collaboration, K.~Akiyama {\em et~al.},
  ``{First Sagittarius A* Event Horizon Telescope Results. I. The Shadow of the
  Supermassive Black Hole in the Center of the Milky Way},''
  \href{http://dx.doi.org/10.3847/2041-8213/ac6674}{{\em Astrophys. J. Lett.}
  {\bfseries 930} no.~2, (2022) L12},
  \href{http://arxiv.org/abs/2311.08680}{{\ttfamily arXiv:2311.08680
  [astro-ph.HE]}}.

\bibitem{EventHorizonTelescope:2022wok}
{\bfseries Event Horizon Telescope} Collaboration, K.~Akiyama {\em et~al.},
  ``{First Sagittarius A* Event Horizon Telescope Results. III. Imaging of the
  Galactic Center Supermassive Black Hole},''
  \href{http://dx.doi.org/10.3847/2041-8213/ac6429}{{\em Astrophys. J. Lett.}
  {\bfseries 930} no.~2, (2022) L14},
  \href{http://arxiv.org/abs/2311.09479}{{\ttfamily arXiv:2311.09479
  [astro-ph.HE]}}.

\bibitem{Misner:1972kx}
C.~W. Misner, ``{Interpretation of gravitational-wave observations},''
  \href{http://dx.doi.org/10.1103/PhysRevLett.28.994}{{\em Phys. Rev. Lett.}
  {\bfseries 28} (1972) 994--997}.

\bibitem{Penrose:1969pc}
R.~Penrose, ``{Gravitational collapse: The role of general relativity},''
  \href{http://dx.doi.org/10.1023/A:1016578408204}{{\em Riv. Nuovo Cim.}
  {\bfseries 1} (1969) 252--276}.

\bibitem{Penrose:1971uk}
R.~Penrose and R.~M. Floyd, ``{Extraction of rotational energy from a black
  hole},'' \href{http://dx.doi.org/10.1038/physci229177a0}{{\em Nature}
  {\bfseries 229} (1971) 177--179}.

\bibitem{Christodoulou:1970wf}
D.~Christodoulou, ``{Reversible and irreversible transforations in black hole
  physics},'' \href{http://dx.doi.org/10.1103/PhysRevLett.25.1596}{{\em Phys.
  Rev. Lett.} {\bfseries 25} (1970) 1596--1597}.

\bibitem{Bardeen:1972fi}
J.~M. Bardeen, W.~H. Press, and S.~A. Teukolsky, ``{Rotating black holes:
  Locally nonrotating frames, energy extraction, and scalar synchrotron
  radiation},'' \href{http://dx.doi.org/10.1086/151796}{{\em Astrophys. J.}
  {\bfseries 178} (1972) 347}.

\bibitem{Wald:1974kya}
R.~M. Wald, ``{Energy Limits on the Penrose Process},''
  \href{http://dx.doi.org/10.1086/152959}{{\em Astrophys. J.} {\bfseries 191}
  (1974) 231}.

\bibitem{osti_4222462}
T.~Piran, J.~Shaham, and J.~Katz, ``High efficiency of the penrose mechanism
  for particle collisions,'' \href{http://dx.doi.org/10.1086/181755}{{\em
  Astrophys. J., Lett., v. 196, no. 3, pp. L107-L108} (3, 1975) }.
  \url{https://www.osti.gov/biblio/4222462}.

\bibitem{Teukolsky:1974yv}
S.~A. Teukolsky and W.~H. Press, ``{Perturbations of a rotating black hole. III
  - Interaction of the hole with gravitational and electromagnet ic
  radiation},'' \href{http://dx.doi.org/10.1086/153180}{{\em Astrophys. J.}
  {\bfseries 193} (1974) 443--461}.

\bibitem{Blandford:1977ds}
R.~D. Blandford and R.~L. Znajek, ``{Electromagnetic extractions of energy from
  Kerr black holes},'' \href{http://dx.doi.org/10.1093/mnras/179.3.433}{{\em
  Mon. Not. Roy. Astron. Soc.} {\bfseries 179} (1977) 433--456}.

\bibitem{Takahashi:1990bv}
M.~Takahashi, S.~Nitta, Y.~Tatematsu, and A.~Tomimatsu, ``{MHD flows in Kerr
  geometry: Energy extraction from black holes},''.

\bibitem{McKinney:2004ka}
J.~C. McKinney and C.~F. Gammie, ``{A Measurement of the electromagnetic
  luminosity of a Kerr black hole},''
  \href{http://dx.doi.org/10.1086/422244}{{\em Astrophys. J.} {\bfseries 611}
  (2004) 977--995}, \href{http://arxiv.org/abs/astro-ph/0404512}{{\ttfamily
  arXiv:astro-ph/0404512}}.

\bibitem{Hawley:2005xs}
J.~F. Hawley and J.~H. Krolik, ``{Magnetically driven jets in the kerr
  metric},'' \href{http://dx.doi.org/10.1086/500385}{{\em Astrophys. J.}
  {\bfseries 641} (2006) 103--116},
  \href{http://arxiv.org/abs/astro-ph/0512227}{{\ttfamily
  arXiv:astro-ph/0512227}}.

\bibitem{Komissarov:2007rc}
S.~S. Komissarov and J.~C. McKinney, ``{Meissner effect and Blandford-Znajek
  mechanism in conductive black hole magnetospheres},''
  \href{http://dx.doi.org/10.1111/j.1745-3933.2007.00301.x}{{\em Mon. Not. Roy.
  Astron. Soc.} {\bfseries 377} (2007) L49--L53},
  \href{http://arxiv.org/abs/astro-ph/0702269}{{\ttfamily
  arXiv:astro-ph/0702269}}.

\bibitem{Tchekhovskoy:2011zx}
A.~Tchekhovskoy, R.~Narayan, and J.~C. McKinney, ``{Efficient Generation of
  Jets from Magnetically Arrested Accretion on a Rapidly Spinning Black
  Hole},'' \href{http://dx.doi.org/10.1111/j.1745-3933.2011.01147.x}{{\em Mon.
  Not. Roy. Astron. Soc.} {\bfseries 418} (2011) L79--L83},
  \href{http://arxiv.org/abs/1108.0412}{{\ttfamily arXiv:1108.0412
  [astro-ph.HE]}}.

\bibitem{Lee:1999se}
H.~K. Lee, R.~A. M.~J. Wijers, and G.~E. Brown, ``{The Blandford-Znajek process
  as a central engine for a gamma-ray burst},''
  \href{http://dx.doi.org/10.1016/S0370-1573(99)00084-8}{{\em Phys. Rept.}
  {\bfseries 325} (2000) 83--114},
  \href{http://arxiv.org/abs/astro-ph/9906213}{{\ttfamily
  arXiv:astro-ph/9906213}}.

\bibitem{Tchekhovskoy:2008gq}
A.~Tchekhovskoy, J.~C. McKinney, and R.~Narayan, ``{Simulations of
  Ultrarelativistic Magnetodynamic Jets from Gamma-ray Burst Engines},''
  \href{http://dx.doi.org/10.1111/j.1365-2966.2008.13425.x}{{\em Mon. Not. Roy.
  Astron. Soc.} {\bfseries 388} (2008) 551},
  \href{http://arxiv.org/abs/0803.3807}{{\ttfamily arXiv:0803.3807
  [astro-ph]}}.

\bibitem{Komissarov:2009dn}
S.~S. Komissarov and M.~V. Barkov, ``{Activation of the Blandford-Znajek
  mechanism in collapsing stars},''
  \href{http://dx.doi.org/10.1111/j.1365-2966.2009.14831.x}{{\em Mon. Not. Roy.
  Astron. Soc.} {\bfseries 397} (2009) 1153},
  \href{http://arxiv.org/abs/0902.2881}{{\ttfamily arXiv:0902.2881
  [astro-ph.HE]}}.

\bibitem{Aharonian:2007ig}
F.~Aharonian {\em et~al.}, ``{An Exceptional Very High Energy Gamma-Ray Flare
  of PKS 2155-304},'' \href{http://dx.doi.org/10.1086/520635}{{\em Astrophys.
  J. Lett.} {\bfseries 664} (2007) L71--L78},
  \href{http://arxiv.org/abs/0706.0797}{{\ttfamily arXiv:0706.0797
  [astro-ph]}}.

\bibitem{HESS:2009cfm}
{\bfseries H.E.S.S.} Collaboration, F.~Aharonian {\em et~al.}, ``{Discovery of
  very high energy gamma-ray emission from Centaurus A with H.E.S.S},''
  \href{http://dx.doi.org/10.1088/0004-637X/695/1/L40}{{\em Astrophys. J.
  Lett.} {\bfseries 695} (2009) L40--L44},
  \href{http://arxiv.org/abs/0903.1582}{{\ttfamily arXiv:0903.1582
  [astro-ph.CO]}}.

\bibitem{Albert:2007zd}
J.~Albert {\em et~al.}, ``{Variable VHE gamma-ray emission from Markarian
  501},'' \href{http://dx.doi.org/10.1086/521382}{{\em Astrophys. J.}
  {\bfseries 669} (2007) 862--883},
  \href{http://arxiv.org/abs/astro-ph/0702008}{{\ttfamily
  arXiv:astro-ph/0702008}}.

\bibitem{Aleksic:2014xsg}
J.~Aleksic {\em et~al.}, ``{Black hole lightning due to particle acceleration
  at subhorizon scales},''
  \href{http://dx.doi.org/10.1126/science.1256183}{{\em Science} {\bfseries
  346} (2014) 1080--1084}, \href{http://arxiv.org/abs/1412.4936}{{\ttfamily
  arXiv:1412.4936 [astro-ph.HE]}}.

\bibitem{VERITAS:2010udc}
{\bfseries VERITAS} Collaboration, V.~A. Acciari {\em et~al.}, ``{VERITAS 2008
  - 2009 monitoring of the variable gamma-ray source M87},''
  \href{http://dx.doi.org/10.1088/0004-637X/716/1/819}{{\em Astrophys. J.}
  {\bfseries 716} (2010) 819--824},
  \href{http://arxiv.org/abs/1005.0367}{{\ttfamily arXiv:1005.0367
  [astro-ph.CO]}}.

\bibitem{Aliu:2011xm}
E.~Aliu {\em et~al.}, ``{VERITAS Observations of day-scale flaring of M87 in
  2010 April},'' \href{http://dx.doi.org/10.1088/0004-637X/746/2/141}{{\em
  Astrophys. J.} {\bfseries 746} (2012) 141},
  \href{http://arxiv.org/abs/1112.4518}{{\ttfamily arXiv:1112.4518
  [astro-ph.CO]}}.

\bibitem{Baganoff:2001kw}
F.~K. Baganoff {\em et~al.}, ``{Rapid X-ray flaring from the direction of the
  supermassive black hole at the galactic centre},''
  \href{http://dx.doi.org/10.1038/35092510}{{\em Nature} {\bfseries 413} (2001)
  45--48}, \href{http://arxiv.org/abs/astro-ph/0109367}{{\ttfamily
  arXiv:astro-ph/0109367}}.

\bibitem{Eckart:2004ka}
A.~Eckart {\em et~al.}, ``{First simultaneous NIR / x-ray detection of a flare
  from Sgr A*},'' \href{http://dx.doi.org/10.1051/0004-6361:20040495}{{\em
  Astron. Astrophys.} {\bfseries 427} (2004) 1--11},
  \href{http://arxiv.org/abs/astro-ph/0403577}{{\ttfamily
  arXiv:astro-ph/0403577}}.

\bibitem{Neilsen:2014kva}
J.~Neilsen {\em et~al.}, ``{The X-ray Flux Distribution of Sagittarius A* as
  Seen by Chandra},'' \href{http://dx.doi.org/10.1088/0004-637X/799/2/199}{{\em
  Astrophys. J.} {\bfseries 799} no.~2, (2015) 199},
  \href{http://arxiv.org/abs/1412.3106}{{\ttfamily arXiv:1412.3106
  [astro-ph.HE]}}.

\bibitem{GRAVITY:2021hxs}
{\bfseries GRAVITY} Collaboration, R.~Abuter {\em et~al.}, ``{Constraining
  particle acceleration in Sgr A* with simultaneous GRAVITY, Spitzer, NuSTAR,
  and Chandra observations},''
  \href{http://dx.doi.org/10.1051/0004-6361/202140981}{{\em Astron. Astrophys.}
  {\bfseries 654} (2021) A22},
  \href{http://arxiv.org/abs/2107.01096}{{\ttfamily arXiv:2107.01096
  [astro-ph.HE]}}.

\bibitem{Koide:2008xr}
S.~Koide and K.~Arai, ``{Energy Extraction from a Rotating Black Hole by
  Magnetic Reconnection in Ergosphere},''
  \href{http://dx.doi.org/10.1086/589497}{{\em Astrophys. J.} {\bfseries 682}
  (2008) 1124}, \href{http://arxiv.org/abs/0805.0044}{{\ttfamily
  arXiv:0805.0044 [astro-ph]}}.

\bibitem{Comisso:2016pyg}
L.~Comisso, M.~Lingam, Y.~M. Huang, and A.~Bhattacharjee, ``{General Theory of
  the Plasmoid Instability},'' \href{http://dx.doi.org/10.1063/1.4964481}{{\em
  Phys. Plasmas} {\bfseries 23} (2016) 100702},
  \href{http://arxiv.org/abs/1608.04692}{{\ttfamily arXiv:1608.04692
  [physics.plasm-ph]}}.

\bibitem{Comisso:2017arh}
L.~Comisso, M.~Lingam, Y.-M. Huang, and A.~Bhattacharjee, ``{Plasmoid
  Instability in Forming Current Sheets},''
  \href{http://dx.doi.org/10.3847/1538-4357/aa9789}{{\em Astrophys. J.}
  {\bfseries 850} no.~2, (2017) 142},
  \href{http://arxiv.org/abs/1707.01862}{{\ttfamily arXiv:1707.01862
  [astro-ph.HE]}}.

\bibitem{PhysRevLett.103.065004}
W.~Daughton, V.~Roytershteyn, B.~J. Albright, H.~Karimabadi, L.~Yin, and K.~J.
  Bowers, ``Transition from collisional to kinetic regimes in large-scale
  reconnection layers,''
  \href{http://dx.doi.org/10.1103/PhysRevLett.103.065004}{{\em Phys. Rev.
  Lett.} {\bfseries 103} (Aug, 2009) 065004}.
  \url{https://link.aps.org/doi/10.1103/PhysRevLett.103.065004}.

\bibitem{Uzdensky:2014uda}
D.~A. Uzdensky and N.~F. Loureiro, ``{Magnetic Reconnection Onset via
  Disruption of a Forming Current Sheet by the Tearing Instability},''
  \href{http://dx.doi.org/10.1103/PhysRevLett.116.105003}{{\em Phys. Rev.
  Lett.} {\bfseries 116} no.~10, (2016) 105003},
  \href{http://arxiv.org/abs/1411.4295}{{\ttfamily arXiv:1411.4295
  [astro-ph.SR]}}.

\bibitem{Bhattacharjee2009}
A.~Bhattacharjee, Y.-M. Huang, H.~Yang, and B.~Rogers, ``{Fast reconnection in
  high-Lundquist-number plasmas due to the plasmoid Instability},''
  \href{http://dx.doi.org/10.1063/1.3264103}{{\em Physics of Plasmas}
  {\bfseries 16} no.~11, (11, 2009) 112102},
  \href{http://arxiv.org/abs/https://pubs.aip.org/aip/pop/article-pdf/doi/10.1063/1.3264103/15788011/112102\_1\_online.pdf}{{\ttfamily
  https://pubs.aip.org/aip/pop/article-pdf/doi/10.1063/1.3264103/15788011/112102\_1\_online.pdf}}.
  \url{https://doi.org/10.1063/1.3264103}.

\bibitem{Eatough:2013nva}
R.~P. Eatough {\em et~al.}, ``{A strong magnetic field around the supermassive
  black hole at the centre of the Galaxy},''
  \href{http://dx.doi.org/10.1038/nature12499}{{\em Nature} {\bfseries 501}
  (2013) 391--394}, \href{http://arxiv.org/abs/1308.3147}{{\ttfamily
  arXiv:1308.3147 [astro-ph.GA]}}.

\bibitem{Ripperda:2021zpn}
B.~Ripperda, M.~Liska, K.~Chatterjee, G.~Musoke, A.~A. Philippov, S.~B.
  Markoff, A.~Tchekhovskoy, and Z.~Younsi, ``{Black Hole Flares: Ejection of
  Accreted Magnetic Flux through 3D Plasmoid-mediated Reconnection},''
  \href{http://dx.doi.org/10.3847/2041-8213/ac46a1}{{\em Astrophys. J. Lett.}
  {\bfseries 924} no.~2, (2022) L32},
  \href{http://arxiv.org/abs/2109.15115}{{\ttfamily arXiv:2109.15115
  [astro-ph.HE]}}.

\bibitem{Ripperda:2020bpz}
B.~Ripperda, F.~Bacchini, and A.~Philippov, ``{Magnetic Reconnection and Hot
  Spot Formation in Black Hole Accretion Disks},''
  \href{http://dx.doi.org/10.3847/1538-4357/ababab}{{\em Astrophys. J.}
  {\bfseries 900} no.~2, (2020) 100},
  \href{http://arxiv.org/abs/2003.04330}{{\ttfamily arXiv:2003.04330
  [astro-ph.HE]}}.

\bibitem{Parfrey:2018dnc}
K.~Parfrey, A.~Philippov, and B.~Cerutti, ``{First-Principles Plasma
  Simulations of Black-Hole Jet Launching},''
  \href{http://dx.doi.org/10.1103/PhysRevLett.122.035101}{{\em Phys. Rev.
  Lett.} {\bfseries 122} no.~3, (2019) 035101},
  \href{http://arxiv.org/abs/1810.03613}{{\ttfamily arXiv:1810.03613
  [astro-ph.HE]}}.

\bibitem{Komissarov:2005wj}
S.~S. Komissarov, ``{Observations of the Blandford-Znajek and the MHD Penrose
  processes in computer simulations of black hole magnetospheres},''
  \href{http://dx.doi.org/10.1111/j.1365-2966.2005.08974.x}{{\em Mon. Not. Roy.
  Astron. Soc.} {\bfseries 359} (2005) 801--808},
  \href{http://arxiv.org/abs/astro-ph/0501599}{{\ttfamily
  arXiv:astro-ph/0501599}}.

\bibitem{East:2018ayf}
W.~E. East and H.~Yang, ``{Magnetosphere of a spinning black hole and the role
  of the current sheet},''
  \href{http://dx.doi.org/10.1103/PhysRevD.98.023008}{{\em Phys. Rev. D}
  {\bfseries 98} no.~2, (2018) 023008},
  \href{http://arxiv.org/abs/1805.05952}{{\ttfamily arXiv:1805.05952
  [astro-ph.HE]}}.

\bibitem{Comisso:2020ykg}
L.~Comisso and F.~A. Asenjo, ``{Magnetic Reconnection as a Mechanism for Energy
  Extraction from Rotating Black Holes},''
  \href{http://dx.doi.org/10.1103/PhysRevD.103.023014}{{\em Phys. Rev. D}
  {\bfseries 103} no.~2, (2021) 023014},
  \href{http://arxiv.org/abs/2012.00879}{{\ttfamily arXiv:2012.00879
  [astro-ph.HE]}}.

\bibitem{Wang:2022qmg}
C.-H. Wang, C.-Q. Pang, and S.-W. Wei, ``{Extracting energy via magnetic
  reconnection from Kerr\textendash{}de Sitter black holes},''
  \href{http://dx.doi.org/10.1103/PhysRevD.106.124050}{{\em Phys. Rev. D}
  {\bfseries 106} no.~12, (2022) 124050},
  \href{http://arxiv.org/abs/2209.08837}{{\ttfamily arXiv:2209.08837 [gr-qc]}}.

\bibitem{Liu:2022qnr}
W.~Liu, ``{Energy Extraction via Magnetic Reconnection in the Ergosphere of a
  Rotating Non-Kerr Black Hole},''
  \href{http://dx.doi.org/10.3847/1538-4357/ac3de3}{{\em Astrophys. J.}
  {\bfseries 925} no.~2, (2022) 149},
  \href{http://arxiv.org/abs/2204.07338}{{\ttfamily arXiv:2204.07338
  [astro-ph.HE]}}.

\bibitem{Li:2023htz}
Z.~Li and F.~Yuan, ``{Energy extraction via Comisso-Asenjo mechanism from
  rotating hairy black hole},''
  \href{http://dx.doi.org/10.1103/PhysRevD.108.024039}{{\em Phys. Rev. D}
  {\bfseries 108} no.~2, (2023) 024039},
  \href{http://arxiv.org/abs/2304.12553}{{\ttfamily arXiv:2304.12553 [gr-qc]}}.

\bibitem{Li:2023nmy}
Z.~Li, X.-K. Guo, and F.~Yuan, ``{Energy extraction from rotating regular black
  hole via Comisso-Asenjo mechanism},''
  \href{http://dx.doi.org/10.1103/PhysRevD.108.044067}{{\em Phys. Rev. D}
  {\bfseries 108} no.~4, (2023) 044067},
  \href{http://arxiv.org/abs/2304.08831}{{\ttfamily arXiv:2304.08831 [gr-qc]}}.

\bibitem{Zhang:2024rvk}
S.-J. Zhang, ``{Energy extraction via magnetic reconnection in
  Konoplya-Rezzolla-Zhidenko parametrized black holes},''
  \href{http://dx.doi.org/10.1103/PhysRevD.109.084066}{{\em Phys. Rev. D}
  {\bfseries 109} no.~8, (2024) 084066},
  \href{http://arxiv.org/abs/2402.15050}{{\ttfamily arXiv:2402.15050 [gr-qc]}}.

\bibitem{Khodadi:2022dff}
M.~Khodadi, ``{Magnetic reconnection and energy extraction from a spinning
  black hole with broken Lorentz symmetry},''
  \href{http://dx.doi.org/10.1103/PhysRevD.105.023025}{{\em Phys. Rev. D}
  {\bfseries 105} no.~2, (2022) 023025},
  \href{http://arxiv.org/abs/2201.02765}{{\ttfamily arXiv:2201.02765 [gr-qc]}}.

\bibitem{Khodadi:2023juk}
M.~Khodadi, D.~F. Mota, and A.~Sheykhi, ``{Harvesting energy driven by
  Comisso-Asenjo process from Kerr-MOG black holes},''
  \href{http://dx.doi.org/10.1088/1475-7516/2023/10/034}{{\em JCAP} {\bfseries
  10} (2023) 034}, \href{http://arxiv.org/abs/2307.00478}{{\ttfamily
  arXiv:2307.00478 [astro-ph.HE]}}.

\bibitem{Zhang:2024ptp}
S.-J. Zhang, ``{Energy extraction via magnetic reconnection in magnetized black
  holes},'' \href{http://arxiv.org/abs/2405.16941}{{\ttfamily arXiv:2405.16941
  [gr-qc]}}.

\bibitem{Shaymatov:2023dtt}
S.~Shaymatov, M.~Alloqulov, B.~Ahmedov, and A.~Wang,
  ``{Kerr-Newman-modified-gravity black hole's impact on the magnetic
  reconnection},'' \href{http://arxiv.org/abs/2307.03012}{{\ttfamily
  arXiv:2307.03012 [gr-qc]}}.

\bibitem{Carleo:2022qlv}
A.~Carleo, G.~Lambiase, and L.~Mastrototaro, ``{Energy extraction via magnetic
  reconnection in Lorentz breaking Kerr\textendash{}Sen and Kiselev black
  holes},'' \href{http://dx.doi.org/10.1140/epjc/s10052-022-10751-w}{{\em Eur.
  Phys. J. C} {\bfseries 82} no.~9, (2022) 776},
  \href{http://arxiv.org/abs/2206.12988}{{\ttfamily arXiv:2206.12988 [gr-qc]}}.

\bibitem{Ye:2023xyv}
X.~Ye, C.-H. Wang, and S.-W. Wei, ``{Extracting spinning wormhole energy via
  Comisso-Asenjo process},''
  \href{http://dx.doi.org/10.1088/1475-7516/2023/12/030}{{\em JCAP} {\bfseries
  12} (2023) 030}, \href{http://arxiv.org/abs/2306.12097}{{\ttfamily
  arXiv:2306.12097 [gr-qc]}}.

\bibitem{Chen:2024ggq}
B.~Chen, Y.~Hou, J.~Li, and Y.~Shen, ``{Energy Extraction from a Kerr Black
  Hole via Magnetic Reconnection within the Plunging Region},''
  \href{http://arxiv.org/abs/2405.11488}{{\ttfamily arXiv:2405.11488 [gr-qc]}}.

\bibitem{Hou:2018avu}
X.~Hou, Z.~Xu, and J.~Wang, ``{Rotating Black Hole Shadow in Perfect Fluid Dark
  Matter},'' \href{http://dx.doi.org/10.1088/1475-7516/2018/12/040}{{\em JCAP}
  {\bfseries 12} (2018) 040}, \href{http://arxiv.org/abs/1810.06381}{{\ttfamily
  arXiv:1810.06381 [gr-qc]}}.

\bibitem{Haroon:2018ryd}
S.~Haroon, M.~Jamil, K.~Jusufi, K.~Lin, and R.~B. Mann, ``{Shadow and
  Deflection Angle of Rotating Black Holes in Perfect Fluid Dark Matter with a
  Cosmological Constant},''
  \href{http://dx.doi.org/10.1103/PhysRevD.99.044015}{{\em Phys. Rev. D}
  {\bfseries 99} no.~4, (2019) 044015},
  \href{http://arxiv.org/abs/1810.04103}{{\ttfamily arXiv:1810.04103 [gr-qc]}}.

\bibitem{Rubin:1980zd}
V.~C. Rubin, N.~Thonnard, and W.~K. Ford, Jr., ``{Rotational properties of 21
  SC galaxies with a large range of luminosities and radii, from NGC 4605 /R =
  4kpc/ to UGC 2885 /R = 122 kpc/},''
  \href{http://dx.doi.org/10.1086/158003}{{\em Astrophys. J.} {\bfseries 238}
  (1980) 471}.

\bibitem{Persic:1995ru}
M.~Persic, P.~Salucci, and F.~Stel, ``{The Universal rotation curve of spiral
  galaxies: 1. The Dark matter connection},''
  \href{http://dx.doi.org/10.1093/mnras/278.1.27}{{\em Mon. Not. Roy. Astron.
  Soc.} {\bfseries 281} (1996) 27},
  \href{http://arxiv.org/abs/astro-ph/9506004}{{\ttfamily
  arXiv:astro-ph/9506004}}.

\bibitem{Bertone:2016nfn}
G.~Bertone and D.~Hooper, ``{History of dark matter},''
  \href{http://dx.doi.org/10.1103/RevModPhys.90.045002}{{\em Rev. Mod. Phys.}
  {\bfseries 90} no.~4, (2018) 045002},
  \href{http://arxiv.org/abs/1605.04909}{{\ttfamily arXiv:1605.04909
  [astro-ph.CO]}}.

\bibitem{Jusufi:2019nrn}
K.~Jusufi, M.~Jamil, P.~Salucci, T.~Zhu, and S.~Haroon, ``{Black Hole
  Surrounded by a Dark Matter Halo in the M87 Galactic Center and its
  Identification with Shadow Images},''
  \href{http://dx.doi.org/10.1103/PhysRevD.100.044012}{{\em Phys. Rev. D}
  {\bfseries 100} no.~4, (2019) 044012},
  \href{http://arxiv.org/abs/1905.11803}{{\ttfamily arXiv:1905.11803
  [physics.gen-ph]}}.

\bibitem{Kiselev:2004vy}
V.~V. Kiselev, ``{Vector field and rotational curves in dark galactic halos},''
  \href{http://dx.doi.org/10.1088/0264-9381/22/3/007}{{\em Class. Quant. Grav.}
  {\bfseries 22} (2005) 541--558},
  \href{http://arxiv.org/abs/gr-qc/0404042}{{\ttfamily arXiv:gr-qc/0404042}}.

\bibitem{Navarro:1995iw}
J.~F. Navarro, C.~S. Frenk, and S.~D.~M. White, ``{The Structure of cold dark
  matter halos},'' \href{http://dx.doi.org/10.1086/177173}{{\em Astrophys. J.}
  {\bfseries 462} (1996) 563--575},
  \href{http://arxiv.org/abs/astro-ph/9508025}{{\ttfamily
  arXiv:astro-ph/9508025}}.

\bibitem{Kiselev:2002dx}
V.~V. Kiselev, ``{Quintessence and black holes},''
  \href{http://dx.doi.org/10.1088/0264-9381/20/6/310}{{\em Class. Quant. Grav.}
  {\bfseries 20} (2003) 1187--1198},
  \href{http://arxiv.org/abs/gr-qc/0210040}{{\ttfamily arXiv:gr-qc/0210040}}.

\bibitem{Navarro:1996gj}
J.~F. Navarro, C.~S. Frenk, and S.~D.~M. White, ``{A Universal density profile
  from hierarchical clustering},'' \href{http://dx.doi.org/10.1086/304888}{{\em
  Astrophys. J.} {\bfseries 490} (1997) 493--508},
  \href{http://arxiv.org/abs/astro-ph/9611107}{{\ttfamily
  arXiv:astro-ph/9611107}}.

\bibitem{Burkert:1995yz}
A.~Burkert, ``{The Structure of dark matter halos in dwarf galaxies},''
  \href{http://dx.doi.org/10.1086/309560}{{\em Astrophys. J. Lett.} {\bfseries
  447} (1995) L25}, \href{http://arxiv.org/abs/astro-ph/9504041}{{\ttfamily
  arXiv:astro-ph/9504041}}.

\bibitem{Wang:2019ftp}
J.~Wang, S.~Bose, C.~S. Frenk, L.~Gao, A.~Jenkins, V.~Springel, and S.~D.~M.
  White, ``{Universal structure of dark matter haloes over a mass range of 20
  orders of magnitude},''
  \href{http://dx.doi.org/10.1038/s41586-020-2642-9}{{\em Nature} {\bfseries
  585} no.~7823, (2020) 39--42},
  \href{http://arxiv.org/abs/1911.09720}{{\ttfamily arXiv:1911.09720
  [astro-ph.CO]}}.

\bibitem{Xu:2018wow}
Z.~Xu, X.~Hou, X.~Gong, and J.~Wang, ``{Black Hole Space-time In Dark Matter
  Halo},'' \href{http://dx.doi.org/10.1088/1475-7516/2018/09/038}{{\em JCAP}
  {\bfseries 09} (2018) 038}, \href{http://arxiv.org/abs/1803.00767}{{\ttfamily
  arXiv:1803.00767 [gr-qc]}}.

\bibitem{Xu:2020jpv}
Z.~Xu, X.~Gong, and S.-N. Zhang, ``{Black hole immersed dark matter halo},''
  \href{http://dx.doi.org/10.1103/PhysRevD.101.024029}{{\em Phys. Rev. D}
  {\bfseries 101} no.~2, (2020) 024029}.

\bibitem{Kiselev:2003ah}
V.~V. Kiselev, ``{Quintessential solution of dark matter rotation curves and
  its simulation by extra dimensions},''
  \href{http://arxiv.org/abs/gr-qc/0303031}{{\ttfamily arXiv:gr-qc/0303031}}.

\bibitem{Rahaman:2010xs}
F.~Rahaman, K.~K. Nandi, A.~Bhadra, M.~Kalam, and K.~Chakraborty, ``{Perfect
  Fluid Dark Matter},''
  \href{http://dx.doi.org/10.1016/j.physletb.2010.09.038}{{\em Phys. Lett. B}
  {\bfseries 694} (2011) 10--15},
  \href{http://arxiv.org/abs/1009.3572}{{\ttfamily arXiv:1009.3572 [gr-qc]}}.

\bibitem{Li:2012zx}
M.-H. Li and K.-C. Yang, ``{Galactic Dark Matter in the Phantom Field},''
  \href{http://dx.doi.org/10.1103/PhysRevD.86.123015}{{\em Phys. Rev. D}
  {\bfseries 86} (2012) 123015},
  \href{http://arxiv.org/abs/1204.3178}{{\ttfamily arXiv:1204.3178
  [astro-ph.CO]}}.

\bibitem{Heydari-Fard:2022xhr}
M.~Heydari-Fard, S.~G. Honarvar, and M.~Heydari-Fard, ``{Thin accretion disc
  luminosity and its image around rotating black holes in perfect fluid dark
  matter},'' \href{http://dx.doi.org/10.1093/mnras/stad558}{{\em Mon. Not. Roy.
  Astron. Soc.} {\bfseries 521} no.~1, (2023) 708--716},
  \href{http://arxiv.org/abs/2210.04173}{{\ttfamily arXiv:2210.04173 [gr-qc]}}.

\bibitem{Das:2020yxw}
A.~Das, A.~Saha, and S.~Gangopadhyay, ``{Investigation of circular geodesics in
  a rotating charged black hole in the presence of perfect fluid dark
  matter},'' \href{http://dx.doi.org/10.1088/1361-6382/abd95b}{{\em Class.
  Quant. Grav.} {\bfseries 38} no.~6, (2021) 065015},
  \href{http://arxiv.org/abs/2009.03644}{{\ttfamily arXiv:2009.03644 [gr-qc]}}.

\bibitem{Xu:2017bpz}
Z.~Xu, J.~Wang, and X.~Hou, ``{Kerr\textendash{}anti-de Sitter/de Sitter black
  hole in perfect fluid dark matter background},''
  \href{http://dx.doi.org/10.1088/1361-6382/aabcb6}{{\em Class. Quant. Grav.}
  {\bfseries 35} no.~11, (2018) 115003},
  \href{http://arxiv.org/abs/1711.04538}{{\ttfamily arXiv:1711.04538 [gr-qc]}}.

\bibitem{Rizwan:2018rgs}
M.~Rizwan, M.~Jamil, and K.~Jusufi, ``{Distinguishing a Kerr-like black hole
  and a naked singularity in perfect fluid dark matter via precession
  frequencies},'' \href{http://dx.doi.org/10.1103/PhysRevD.99.024050}{{\em
  Phys. Rev. D} {\bfseries 99} no.~2, (2019) 024050},
  \href{http://arxiv.org/abs/1812.01331}{{\ttfamily arXiv:1812.01331 [gr-qc]}}.

\bibitem{Liang:2023jrj}
X.~Liang, Y.-P. Hu, C.-H. Wu, and Y.-S. An, ``{Thermodynamics and evaporation
  of perfect fluid dark matter black hole in phantom background},''
  \href{http://dx.doi.org/10.1140/epjc/s10052-023-12200-8}{{\em Eur. Phys. J.
  C} {\bfseries 83} no.~11, (2023) 1009},
  \href{http://arxiv.org/abs/2308.00308}{{\ttfamily arXiv:2308.00308 [gr-qc]}}.

\bibitem{Rakhimova:2023rie}
G.~Rakhimova, F.~Atamurotov, F.~Javed, A.~Abdujabbarov, and G.~Mustafa,
  ``{Thermodynamical analysis of charged rotating black hole surrounded by
  perfect fluid dark matter},''
  \href{http://dx.doi.org/10.1016/j.nuclphysb.2023.116363}{{\em Nucl. Phys. B}
  {\bfseries 996} (2023) 116363}.

\bibitem{Liu:2024qso}
Q.-X. Liu, Y.-P. Hu, T.-T. Sui, and Y.-S. An, ``{Superradiance of rotating
  black holes surrounded by dark matter},''
  \href{http://arxiv.org/abs/2406.04611}{{\ttfamily arXiv:2406.04611 [gr-qc]}}.

\bibitem{Mannheim:1988dj}
P.~D. Mannheim and D.~Kazanas, ``{Exact Vacuum Solution to Conformal Weyl
  Gravity and Galactic Rotation Curves},''
  \href{http://dx.doi.org/10.1086/167623}{{\em Astrophys. J.} {\bfseries 342}
  (1989) 635--638}.

\end{thebibliography}\endgroup
\biboptions{sort&compress}
%% else use the following coding to input the bibitems directly in the
%% TeX file.

%% Refer following link for more details about bibliography and citations.
%% https://en.wikibooks.org/wiki/LaTeX/Bibliography_Management

%\begin{thebibliography}{00}

%% For numbered reference style
%% \bibitem{label}
%% Text of bibliographic item

%\bibitem{lamport94}
%  Leslie Lamport,
%  \textit{\LaTeX: a document preparation system},
%  Addison Wesley, Massachusetts,
%  2nd edition,
%  1994.

%\end{thebibliography}
\end{document}